\documentclass[10pt]{article}
\usepackage{amsmath,amssymb,graphicx}
\usepackage{accents}
\usepackage{bbm}
\usepackage{lipsum}
\usepackage{array}
\usepackage{multirow}
\usepackage{tabularx,tabulary,ragged2e,booktabs,caption}
\usepackage{here}
\usepackage[toc,page]{appendix}
\usepackage{lscape}
\usepackage{subcaption}
\def \halfbox #1%
{%
{\setbox 0 = \hbox {$\scriptstyle #1$}%
\dimen 0 = \ht 0
\advance \dimen 0 by 1,0 pt
\ht 0 = \dimen 0
\dimen 0 = \dp 0
\advance \dimen 0 by 1,50 pt
\dp 0 = \dimen 0
\vbox {\hrule \hbox {\box 0 \vrule }}%
}%
}

\usepackage[usenames,dvipsnames]{color}

\numberwithin{equation}{section}
\numberwithin{figure}{section}
\numberwithin{table}{section}
\allowdisplaybreaks[4]

\definecolor{c20}{rgb}{0.,0,0.}
\definecolor{c30}{rgb}{1,0.5,0.5}
\definecolor{c40}{rgb}{0,.7,0}
\definecolor{c50}{rgb}{0,0,0}

\def\zE#1{\textcolor{c30}{#1}}
\def\zE#1{#1}

\newtheorem{theo}{Theorem}[section]
\newtheorem{sat}[theo]{Proposition}
\newtheorem{de}[theo]{Definition}
\newtheorem{lem}[theo]{Lemma}
\newtheorem{exxa}[theo]{Example}

\newtheorem{korr}[theo]{Corollary}

\newtheorem{remarks}[theo]{Remarks}

\newcommand{\abs}[1]{\lvert #1 \rvert}

\newcommand{\R}{\mathbb{R}}

\newcommand{\BQN}{\begin{eqnarray}}
\newcommand{\EQN}{\end{eqnarray}}

\newcommand{\BQNY}{\begin{eqnarray*}}
\newcommand{\EQNY}{\end{eqnarray*}}

\newcommand{\BS}{\begin{sat}}
\newcommand{\ES}{\end{sat}}
\newcommand{\BT}{\begin{theo}}
\newcommand{\ET}{\end{theo}}
\newcommand{\BK}{\begin{korr}}
\newcommand{\EK}{\end{korr}}

\newcommand{\BD}{\begin{de}}
\newcommand{\ED}{\end{de}}

\newcommand{\BIT}{\begin{itemize}}
\newcommand{\EIT}{\end{itemize}}
\newcommand{\BDI}{\begin{description}}
\newcommand{\EDI}{\end{description}}

\newcommand{\BRM}{\begin{remarks}}
\newcommand{\ERM}{\end{remarks}}

\newcommand{\BTH}{\begin{theo}}
\newcommand{\ETH}{\end{theo}}
\newcommand{\BPR}{\begin{sat}}
\newcommand{\EPR}{\end{sat}}

\newcommand{\BEX}{\begin{exxa}}
\newcommand{\EEX}{\end{exxa}}

\newcommand{\BC}{\begin{cases}}
\newcommand{\EC}{\end{cases}}
\newcommand{\COM}[1]{}
\newcommand{\BL}{\begin{lem}}
\newcommand{\EL}{\end{lem}}

\definecolor{c20}{rgb}{0,0,0}
\def\green#1{\textcolor{c20}{#1}}
\def\green#1{#1}

\definecolor{c10}{rgb}{1,0,0}
\def\red#1{\textcolor{c10}{#1}}
\def\yT#1{\textcolor{c10}{#1}}
\def\yT#1{#1}
\def\red#1{#1}

\definecolor{c30}{rgb}{0,0,1}
\def\blue#1{\textcolor{c30}{#1}}
\def\eH#1{\textcolor{c30}{#1}}
\def\eH#1{#1}

\begin{document}
\begin{center}
			{\large
	\bf On bivariate lifetime modelling in life insurance applications  }\\
 \vskip 0.4 cm

         \centerline{\large      
        Fran{\c c}ois Dufresne \footnote{Department of Actuarial Science, 	Faculty of Business and Economics, University of Lausanne, UNIL-Dorigny 1015 Lausanne, Switzerland}, 
         Enkelejd Hashorva $^{1}$,
           Gildas Ratovomirija $^{1,}$\footnote {
	Vaudoise Assurances, Place de Milan CP 120, 1001 Lausanne,         Switzerland}
	and Youssouf Toukourou $^{1}$
}

\COM{
Fran{\c c}ois Dufresne, Enkelejd Hashorva, Gildas Ravotomirija and Youssouf Toukourou\\

\vspace{0.5cm}
{\it  Dept. of Actuarial Science, Faculty of Business and Economics\\
  \it University of Lausanne UNIL-Dorigny, CH-1015 Lausanne, \textit{Switzerland} \\
\it francois.dufresne@unil.ch; enkelejd.hashorva@unil.ch\\ 
  \it gildas.ravotomirija@unil.ch; youssouf.toukourou@unil.ch\\
  }
  }
\end{center}

\hrulefill
\\
{\bf Abstract}

\medskip
Insurance and annuity products covering several lives require the modelling of the joint distribution of future lifetimes. In the interest of simplifying calculations, it is common in practice to assume that the future lifetimes among a group of people are independent. However, extensive research over the past decades suggests otherwise. In this paper, a copula approach is used to model the dependence between lifetimes within a married couple \eH{using data from a large Canadian insurance company}. As a novelty, the age difference and the \eH{gender} of the elder partner are introduced as an argument of the dependence parameter. \green{Maximum likelihood techniques are} thus implemented for the parameter estimation. Not only do the results make clear that the correlation decreases with age difference, but also the dependence between the lifetimes is higher when husband is older than wife. A goodness-of-fit procedure is applied in order to assess the validity of the model. Finally, considering several products available on the life insurance  market, the paper concludes with practical illustrations. 

\section{Introduction}
Insurance and annuity products covering several lives require the modelling of the joint distribution of future lifetimes. \eH{Commonly in actuarial practice}\yT{,} \eH{the future lifetimes among a group of people are assumed to be independent. This simplyifying assumption is not supported by real insurance data as demonstrated by numerous investigations.}  Joint life annuities issued to married couples offer a very good illustration of this fact. It is well known that husband and wife tend to be exposed to similar risks as they are likely to have the same living habits. For example, Parkes et al. \cite{parkes1970broken} and Ward \cite{ward1976mortality} have brought to light the increased mortality of widowers, often called the {\it broken heart syndrome}. Many contributions have shown that there could be a significant difference between risk-related quantities, such as risk premiums, evaluated according to dependence or independence assumptions. Denuit and Cornet \cite{denuit1999multilife} have measured the effect of lifetime dependencies on the present \red{value} of a widow pension benefit. Based on the data collected in cemeteries, not only do their estimation results confirm that the mortality risk depends on the marital status, but also show that the amounts of premium are reduced approximately by 10 per cent compared to \red{ model} which assumes independence. According to data from a large Canadian insurance company, Frees et al. \cite{frees1996annuity} \green{have demonstrated} that there is a strong positive dependence between joint lives. Their estimation results indicate that annuity values are reduced by approximately 5 per cent compared to model with independence.
\\Introduced by Sklar \cite{sklar1959fonctions}, copulas have been widely used to model the dependence structure of random vectors. In the particular case of bivariate lifetimes, frailty models can be used to describe the common risk factors between husband and wife. Oakes \cite{oakes1989bivariate} has shown that the bivariate distributions generated by frailty models are a subclass of Archimedean copulas. This makes this particular  copula family very attractive for modelling bivariate lifetimes. 
We refer to  \cite{nelsen2007introduction} for a general introduction to copulas and \green{ \cite{CoriA,  CoriB}}, \eH{for applications of Archimedean copula in risk theory}. 

 The Archimedean copula family has been proved valuable in numerous life insurance applications, \green{see e.g.,  \cite{frees1996annuity, brown1999joint, carriere2000bivariate}. In \cite{luciano2008modelling},} the marginal distributions and the copula are fitted separately and, the results show that the dependence increases with age. 
\\
It is known that the level of association between variables is characterized by the value of the dependence parameter. In this paper, a special attention is paid to this dependence parameter. Youn and Shemyakin \cite{youn1999statistical} have introduced the age difference between spouses as an argument of the dependence parameter of the copula. In addition, the sign of the age difference is of great interest in our model. More precisely, we presume that the \eH{gender} of the older member of the couple has an influence on the level of dependence between lifetimes. In order to confirm our hypothesis, four families of Archimedean copulas are discussed namely, Gumbel, Frank, Clayton and Joe copulas,  all these under a Gompertz distribution assumption for marginals. The parameter estimations are based on the maximum likelihood approach using data from a large Canadian insurance company, the same set of data used by Frees et al. \cite{frees1996annuity}. Following  \cite{joe1996estimation} and  \cite{oakes1989bivariate}, a two-step technique, where marginals and copula are estimated separately, is \zE{applied}. The results make clear that the dependence is higher when husband is older than wife.
\\Once the marginal and copula parameters are estimated, one needs to assess the goodness of fit of the model. For example, the likelihood ratio test is used in  \cite{carriere2000bivariate} whereas the model of Youn and Shemyakin \cite{youn1999statistical} is based on the Akaike Information Criterion \zE{(AIC)}. In this paper, following  \cite{gribkova2015non} and  \cite{lawless2011statistical}, we implement a whole goodness of fit procedure to validate the model. Based on the Cram\`er-von Mises statistics, the Gumbel copula, whose dependence parameter is \zE{a} function of the age difference and its sign \eH{gives the best fit.} 
\\The rest of the paper is organized as follows. Section \ref{sec:motivation} \zE{discusses} the main characteristics of the dataset and provides some key facts that motivate our study. Section 3 describes the maximum likelihood procedure used to estimate the marginal distributions. \red{ The dependence models are examined 
in Section 4}. In a first hand, we describe the copula models whose parameter are estimated. Secondly, a bootstrap algorithm is proposed for assessing the goodness of fit of the model. Considering several products available on the life insurance market, numerical applications with real data, including \green{best estimate of liabilities, risk capital and stop loss premiums} are presented in \red{ Section} 5. Section \ref{sec: conclusion} concludes the paper.
\section{Motivation} \label{sec:motivation}
\zE{As already shown in \green{\cite{maeder1996construction}}}, being in a married couple can significantly influence the mortality. Moreover, the remaining lifetimes of male and female in the couple are dependent, \green{see e.g., \cite{carriere2000bivariate, frees1996annuity}.} In this \eH{contribution}, \eH{we aim at modelling the dependence} between the lifetimes of a man and a woman within a married couple. 
\eH{Common dependence measures, which will be used in our} \yT{\COM{model}study}\eH{, are:} 
the Pearson's correlation coefficient $r$, the Kendall's Tau $\tau$, and the \green{ Spearman's Rho $\rho$}. 
\COM{The first one measures the linear relationship between two variables whereas the last two measure the extent to which, as one variable increases, the other variable tends to increase, without requiring that increase to be represented by a linear relationship.\\}
In order to develop these aspects, data
 \footnote{ we wish to thank the Society of Actuaries, through the courtesy of Edward (Jed) Frees and Emiliano Valdez, for allowing the use of the data in this paper.} 
 from a large Canadian life insurance company are used. The dataset contains information from policies that were in force during the observation period, i.e. from December 29, 1988 to December 31, 1993. 
\COM{ Information are relative to the time the annuitants enter the study, the \eH{gender}, the age, the mortality and the annuity amount. }
 Thus, we have $14'947$ contracts among which $14'889$ couples (one male and one female) and the remaining $58$ are contracts where annuitants are both male (22 pairs) or both female (36 pairs). The same dataset has been \eH{analysed} in \green{  \cite{frees1996annuity, carriere2000bivariate, youn1999statistical, gribkova2015non} } among others, also in the framework of modelling bivariate lifetime. Since we are interested in the dependence within the couple, we focus our attention on the male-female contracts.
\\We refer the readers to  \cite{frees1996annuity} for the data processing procedure. The dataset is left truncated as the annuitant informations are recorded only from the date they enter the study; this means that insured who have died before the beginning of the observation period were not taken into account in the study. The dataset is also right censored in the sense that most of the insured were alive at the end of the study. Considering our sample as described above, some couples having several contracts could appear many times. By considering each couple only once, our dataset \eH{consists of} $12'856$ different couples for which, we can draw the following informations:
\begin{itemize}
\item the entry ages $x_m$ and $x_f$ for male and female, respectively,
\item the lifetimes under the observation period $t_m$ and $t_f$ for male and female, respectively, and
\item the binary right censoring indicator $\delta_m$ and $\delta_f$ for male and female, respectively,
\item the couple’s benefit in Canadian Dollar  (CAD)  amount  within a last survivor contract.
\end{itemize}
The entry age is the age at which, the annuitant enters the study. The lifetime at entry age corresponds to the lapse of time during which the individual was alive over the period of study. Therefore, for a male (resp. female) aged $x_m$ (resp. $x_f$) at entry and whose data is not censored i.e. $\delta_m=0$ (resp. $\delta_f=0$), $x_m + t_m$ (resp. $x_f + t_f$) is the age at death. When the data is right censored i.e. $\delta_m=1$ (resp. $\delta_f=1$), the number $x_m +t_m$ (resp. $x_f +t_f $) is the age at the end of the period of study (December 31, 1993). The lifetime is usually equal to $5.055$ years corresponding to the duration of the study period; but it is sometimes less as some people may entry later or die before the end of study. Benefit is paid each year until the death of the last survivor. Its value will be used as an input for the \green{applications} of the model to insurance products in \green{ Section} \ref{subsec: res}. Some summary statistics of the age distribution \red{of} our dataset are displayed in Table \ref{Stat_desc}.
\begin{table}[h]
\centering

\begin{tabular}{|l|r|r|r|r|r|} 
   \hline
     & \multicolumn{2}{c|}{Males age} & \multicolumn{2}{c|}{Females age}  \\
    \cline{2-5}
    \cline{2-5}
     Statistics & Entry & Death & Entry & Death    \\
    \hline
    \hline
    Number & $12'856$ & $1'349$ & $12'856$ & $484$ \\
    \hline
    \hline
    Mean & $67.9$ & $74.41$ & $64.95$ & $73.76$ \\
    Std. dev. & $6.38$ & $7.18$ & $7.26$ & $7.87$\\
    $Median$ & $67.68$ & $74.18$ & $65.27$ & $ 73.09 $  \\
    $10^{th} percentile$ & $60.34$ & $66.00$ & $55.92$ & $64.24$ \\
    $90^{th} percentile$ & $75.41$ & $83.21$ & $73.42$ & $83.92$  \\
    \hline
    \hline
\end{tabular}
\caption{\label{Stat_desc} Summary of the univariate distribution statistics.}
\end {table}
It can be seen that the average entry age is $66.4$ for the entire population, $67.9$ for males and $64.9$ for female; $90\%$ of annuitants are older than $57.9$ at entry and males are older than females by $3$ years on average. Among the $12'856$ couples considered, there are $1349$ males and $484$ females who die during the study period. In addition, there are 11'228 couples where both annuitants are alive at the end of the observation while both spouses are dead for 205 couples.
Based on these $205$ couples, the empirical dependence measures are displayed in the last row of Table \ref{dependence_sign_d}. The values show that the ages at death of spouses are positively correlated.\\
\begin{center}
\begin{tabular}{|l|r|r|r|r|} 
   \hline
    & & \multicolumn{3}{c|}{Dependence measures}\\
    \cline{2-5}
    \cline{2-5}
   & Number   & $r$ & $\rho$ & $\tau$ \\
    \hline
    \hline
    $x_m > x_f$ & $154$ & $0.90$ & $0.88$ & $0.72$ \\
    \hline
    $x_m < x_f$ & $51$ & $0.88$ & $0.86$ & $0.69$ \\
    \hline
    Total & $205$ & $0.82$ & $0.80$ & $0.62$ \\
    \hline
    \hline
\end{tabular}
\captionof{table}{\label{dependence_sign_d} \red{Empirical dependence measures} with respect to the \eH{gender} of the elder partner.}
\end{center}

From the existing literature, \green{ see e.g.,  \cite{denuit1999multilife,  youn1999statistical, denuit2001measuring},} the dependence within a couple is often influenced by three factors:
\begin{itemize}
\item the \textbf{common lifestyle} that husband and wife follow, for example their eating habits,
\item the \textbf{common disaster} that affects simultaneously the husband and his wife, as they are likely to be in the same area when a catastrophic event occurs,
\item the \textbf{broken-heart factor} where the death of one would precipitate the death of the partner, often due to the vacuum caused by the passing away of the companion.
\end{itemize}
Based on the common disaster and the broken-heart, Youn and Shemyakin \cite{youn1999statistical} have introduced the {\it age difference between spouses}. Their results show that the model captures some additional association between lifetime of the spouses that would not be reflected in a model without age difference. It is also observed that, the higher the age difference is, the lower is the dependence. \red{Referring to} the same dataset, Table \ref{dependence_age_difference_d} confirms their results, with $\abs{d}$ the absolute value of $d$ and $d=x_m -x_f$.\\

\begin{center}

\begin{tabular}{|l|r|r|r|r|} 
   \hline
    & & \multicolumn{3}{c|}{Dependence measures}\\
    \cline{2-5}
    \cline{2-5}
   & Number   & $r$ & $\rho$ & $\tau$ \\
    \hline
    \hline
    $0\leq \abs{d} < 2 $ & $83$ & $0.97$ & $0.96$ & $0.84$ \\
    \hline
    $2\leq \abs{d} < 4$ & $50$ & $0.94$ & $0.94$ & $0.82$\\
    \hline
    $ \abs{d} \geq 4$ & $72$ & $0.72$ & $0.63$ & $0.50$ \\
    \hline
    \hline
\end{tabular}
\captionof{table}{\label{dependence_age_difference_d} \red{Empirical dependence measures} with respect to the age difference.}
\end{center}
Our study follows the same lines of idea as these authors. In addition to the age difference, we believe that the \eH{gender} of the elder \green{partner} may have an impact on \red{ their lifetimes dependencies}. Indeed, the fact that the husband is older than the wife may influence their relationship, and indirectly, the dependence factors cited above. The results displayed in Table \ref{dependence_sign_d} clearly show that the spouse lifetime dependencies are higher when \green{$d$} is positive, i.e. when husband is older than wife. The variable {\it \eH{gender} of the elder member} is measured through the sign of the age difference $d$. Table \ref{dependencebyagesex} displays the empirical \green{Kendall's $\tau$} with respect to the age difference and to the \eH{gender} of the elder partner. One can notice that the coefficients can vary for more than $30 \%$ depending on who is the older member of the couple.
\begin{table}[h]
\begin{center}

	\begin{tabular}{|c|c|c|c|c|}
		\hline

		$\tau$ & Total & $0\leq \abs{d}< 2$  &   $ 2  \leq \abs{d}< 4 $ & $\abs{d} \geq 4$   \\

		\hline
		\hline
		$x_m \geq x_f$  & $0.72$ & $0.89 $ & $0.89$ & $0.55$ \\
		\hline
		No. of ($x_m \geq x_f$)  & $154$ & $53 $ &$41$ & $60$\\
		\hline
		$x_m < x_f$  & $0.69$ & $ 0.86$ & $0.86$ & $0.74$\\
		\hline
		No. of ($x_m < x_f$)  & $51$  & $30$ & $9$ & $12$\\
		\hline
		\end{tabular}
		\caption{\label{dependencebyagesex} Kendal'Tau correlation coefficients by age and \eH{gender} of the elder partner.} 
\end{center}
\end{table}
\\In what follows, a bivariate lifetime model will verify our hypothesis. To do this,  marginal distributions for each of the male and female lifetimes are firstly defined and secondly the copula  \red{models are} introduced.
\COM{The copula parameter considered is a function of the absolute value of the age difference $d$ and the \eH{gender} of the elder partner through the sign of $d$.} 
The estimation methods will be detailed in the Section \ref{sec:marg} and Section \ref{sec:copulas_definition}.

\section{Marginal distributions}\label{sec:marg}
\subsection{Background}
The lifetime of a newborn \eH{shall be modelled by} a positive continuous random variable, say $X$ with distribution function (df) $F$ and survival function \green{$S$}. 
The symbol $\left( x\right)$ will be used to denote a live aged $x$ and $T\left( x\right)=\eH{ (X- x) \lvert X> x}$ is the remaining lifetime of $\left( x\right)$. The actuarial symbols $_tp_x$ and $_tq_x$ are, respectively, the survival function and the df of $T\left( x\right) $.
 \COM{
  \green{we have }
\BQNY
_tp_x= \mathbb{P}\left( T\left( x\right) >t\right) =1- {_tq_x}=1-\mathbb{P}\left( T\left( x\right) \leq t\right).
\EQNY
}
\green{Indeed}, the probability, for a live $\left( x \right) $, to remain alive $t$ more years is given by
\BQNY
_tp_x=\mathbb{P}\left( X > x+t\mid X>x\right)=\frac{\mathbb{P}\left( X>x+t\right)}{\mathbb{P}\left( X>x\right)}=\frac{S\left( x+t \right)}{S\left( x\right) }.
\EQNY
\eH{ When $X$ has a probability density function $f$,} 
\COM{
The distribution of $T\left( x\right) $ can also be characterized by its hasard rate function $\mu\left( x\right) $, also called \textit{force of mortality}. This is 
\BQNY
\mu\left( x\right) =\frac{f\left( x\right) }{S\left( x\right)}=-\frac{d \left( \ln S\left( x\right)\right)  }{dx}
\EQNY
where $f\left( x\right)$ is the density function of the lifetime random variable $X$. It describes the mortality density at $x$ given that the live survives until age $x$. The survival function can also be rewritten in terms of the force of mortality as follows
\BQNY
S\left( x\right) =\exp\left( - \int_0^{x}\mu\left( s\right) ds \right).
\EQNY
It follows then that
\begin{equation}
_tp_x=\frac{S\left( x+t\right)}{S\left( x\right)}=\exp\left( - \int_x^{x+t}\mu\left( s\right) ds \right).
\nonumber
\end{equation}
}
 then $T(x)$ 
 has a probability density function given by
\begin{equation}
f_x\left( t\right) = {_tp_x} \; \mu\left( x+t\right).
\nonumber
\end{equation}
where  \red{$\mu(.)$} is the hasard rate function, also called \textit{force of mortality}.\\ 
Several parametric mortality laws such as De Moivre, constant force of mortality, Gompertz, Inverse-Gompertz, Makeham, Gamma, Lognormal and Weibull are used in the literature; see \cite{bowers1986actuarial}. The choice of a specific mortality model is determined mainly by the caracteristics of the available data and the objective of the study. It is well known that the De Moivre law and the constant force of mortality assumptions are interesting for \red{theoretical} purposes whereas Gompertz and Weibull are more appropriate for fitting real data, \green{especially for population of age over $30$}. 
\green{The \green{data set } exploited in this paper 
 regroups essentially policyholders who are at least middle-aged. 
 }
 That is why, in our study, the interest is on the Gompertz law whose caracteristics are defined as follows
\begin{equation}
\mu\left( x\right) =B c^x \;\;\;\;\; \text{ and } \;\;\;\;\;S\left( x\right)=\exp\left( -\frac{B}{\ln c}\left( c^x-1\right) \right) \;\;\;\;\; \text{ with } \;\;\;\;\;B>0, \;\; c>1, \;\; x\geq0.
\nonumber
\end{equation}
\COM{
There is an other reason that motivates the choice of the Gompertz distribution. 
}
\green{In addition},
Frees et al. \cite{frees1996annuity} and Carriere \cite{carriere2000bivariate} have shown that the Gompertz mortality law fits our dataset very well, see \yT{Figure \ref{fig:Gompertz_figure}}.
\COM{
 For reasons detailed in Carriere \cite{carriere1994investigation}, the following reparametrization of the Gompertz law is more convenient for estimation purposes: 
 }
\green{ For estimation purposes the Gompertz law has been reparametrized as follows (see \cite{carriere1994investigation})}
\begin{equation}
e^ {-m/\sigma}=\dfrac{B}{\ln c} \;\;\;\;\;\;\;\;\text{     and    }\;\;\;\;\;\;\;\; e^{1/\sigma}  =c
\nonumber
\end{equation}
from which we obtain
\COM{\green{
\BQN \label{eq:df_Gomperz}
\mu\left( x + t\right) &=&\frac{1}{\sigma} \exp \left(\frac{x+t-m}{\sigma} \right),\notag \\
_{t}p_{x}&=&\exp\left( e^{\frac{x-m}{\sigma}} \left( 1-e^{\frac{t}{\sigma}} \right)  \right), \notag \\
f_x(t)&=&
\frac{1}{\sigma}\exp\left(\frac{x+t_j^{i}-j}{\sigma} \right) 
 \exp\left(  e^{\frac{x-m}{\sigma}}   ( 1-e^{\frac{t}{\sigma}} )  \right), \notag \\
F_x(t) &=& 1- \exp\left( e^{\frac{x-m}{\sigma}} \left( 1-e^{\frac{t}{\sigma}} \right)  \right),
\EQN
}}
\yT{
\BQN \label{eq:df_Gomperz}
\mu\left( x + t\right) &=&\frac{1}{\sigma} \exp \left(\frac{x+t-m}{\sigma} \right),\notag \\
_{t}p_{x}&=&\exp\left( e^{\frac{x-m}{\sigma}} \left( 1-e^{\frac{t}{\sigma}} \right)  \right), \notag \\
f_x(t)&=&
 \exp\left(  e^{\frac{x-m}{\sigma}}   \left( 1-e^{\frac{t}{\sigma}} \right)  \right)
 \frac{1}{\sigma}\exp\left(\frac{x+t-m}{\sigma} \right), \notag \\
F_x(t) &=& 1- \exp\left( e^{\frac{x-m}{\sigma}} \left( 1-e^{\frac{t}{\sigma}} \right)  \right),
\EQN
}
where the mode $m >0$ and the dispersion parameter $\sigma >0$ are the new parameters of the distribution.

\subsection{Maximum likelihood procedure}
In what follows, we will use the \zE{following notation}:
\begin{itemize}
\item The index $j$ indicates the \eH{gender} of the individual, i.e. $j=m$ for male and $j=f$ for female.
\item $\theta_j = \left( m_j, \sigma_j\right) $ denotes the vector of unknown Gompertz parameters for a given \eH{gender} $j$.
\item $n$ is the total number of couples in our data set. Hereafter, a couple means a group of two persons of opposite \eH{gender} that have signed an insurance contract and $i$ is the couple index with $1\leq i \leq n$.
\COM{
The reader should notice that
\begin{equation}
\sum_{i=1}^n \delta_j^{i} = n_j , \;\;\;\;\; j=m,f.
\nonumber
\end{equation}
}
\item For a couple $i$, $t_j^{i}$  is the remaining lifetime observed in the collected data. Indeed, for an individual of \eH{gender} $j$  aged $x_j$, the remaining lifetime $T_j^{i} \left( x\right) $ is a random variable 
\green{such that 
\begin{align}
T_j^{i}\left( x_j\right)&=\min\left( t_j^{i}, B_j^{i} \right) \;\;\;\;\; \text{ and } \;\;\;\;\;  \delta_j^{i} =\mathbf{1}_{\left\lbrace t_j^{i} \geq B_j^{i} \right\rbrace }, \nonumber
\end{align}
\red{where} $B_j^{i}$ is a random censoring point of the individual of \eH{gender} $j$ in the couple $i$.}
\COM{
\item $_tp_{x_j} \left( \theta_j \right)$  is the survival probability for an individual of \eH{gender} $j$ and age $x_j$.
\item $f_{x_j}\left( t^i _j, \theta_j \right)$  denotes, for couple $i$, the probability density of the individual of \eH{gender} $j$ and age $x_j$.
\item $\mu\left( x_j, \theta_j\right)$ denotes the force of mortality at age $x_j$ for \eH{gender} $j$.
}
\end{itemize}
Consider a couple $i$ where the male and female were, respectively, aged $x_m$ and $x_f$ at contract initiation date. 
For each \eH{gender} $j=m,f$, the contribution to the likelihood is given by
\begin{equation} \label{individual_likelihood}
L_j^i\left(  \theta_j \right) = \left[ {_{B_j^{i}}}p_{x_j} \left( \theta_j \right) \right]^{\delta_j^{i}}  \left[ f_{x_j}^i \left( t_j^{i},\theta_j \right)\right]^{1-\delta_j^{i}}.
\end{equation}
\green{
We recall that the dataset is left truncated that is why likelihood function in \eqref{individual_likelihood} has therefore to be conditional on survival to the entry age $x_j$, see e.g., \cite{carriere2000bivariate}.
Therefore, the overall likelihood function can be written as follows
\begin{equation}
 L_j\left( \theta_j \right) =\prod_{i=1}^n L_j^i\left( \theta_j \right),\;\;\;\;\;\;  j=m,f.
\label{eq:marginal_likelihood}
\end{equation}
}

\begin{center}
\begin{figure}[H]
\includegraphics[width=15cm,height=8 cm]{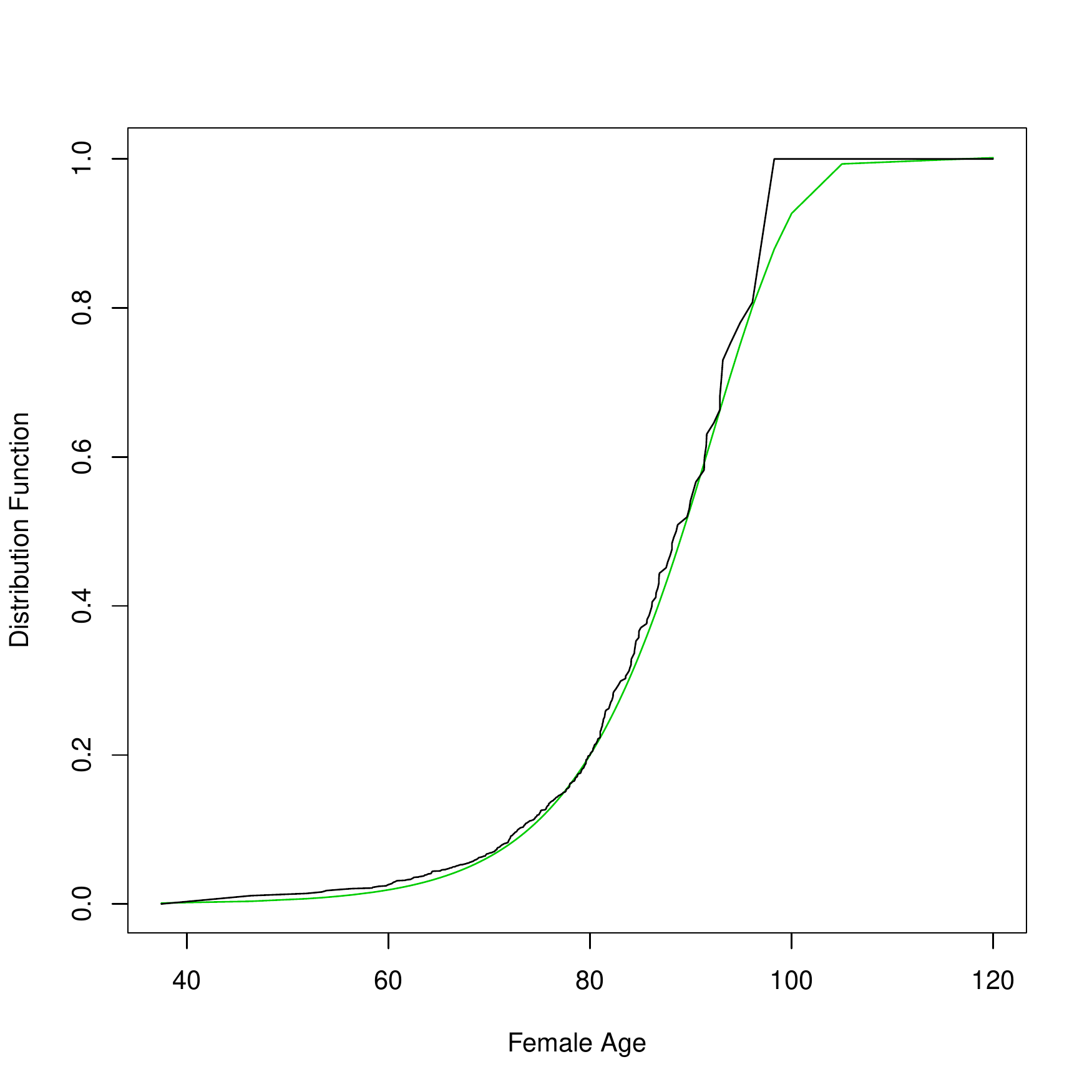}
\caption{Gompertz and Kaplan-Meier fitted female distribution functions} \label{fig:Gompertz_figure}
\end{figure}
\end{center}

By maximizing  the likelihood function in \eqref{eq:marginal_likelihood} using our dataset, the MLE estimates of the Gompertz df are displayed in Table \ref{table: Gompertz_estimates}.
\begin{center}
	\begin{tabular}{|c|c|c|}
		\hline
		
		$\hat{\theta}$ & Estimate & Std. error     \\

		\hline
		$\hat{m}_m$  & 86.378  & 0.289   \\
		\hline
		$\hat{m}_f$  &92.175  &0.527   \\
		\hline
		$\hat{\sigma}_m$  & 9.833  &0.415 \\
		\hline
		$\hat{\sigma}_f$  & 8.114  & 0.392 \\
		\hline
		\end{tabular}
		\captionof{table}{ Gompertz parameter estimates.} \label{table: Gompertz_estimates}
\end{center}
Standard errors are relatively low and estimation shows that the modal age at death is larger for females than for males. This latter can be explained by the fact that women have a longer life expectency than men. A good way to analyse how well the model  performs is to compare with the \textit{Kaplan-Meier (KM) product-limit estimator}  of the dataset. We recall that the KM technique is an approach which consists in estimating non-parametrically the survival function from the empirical data. 
\COM{
We refer the reader to Kaplan and Meier \cite{kaplan1958nonparametric} and the book of Klugman et al. \cite{klugman2012loss} (Chapter 14) to learn more about this estimator. 
}
Figure \ref{fig:Gompertz_figure} compares, for the female group, the KM estimator of the survival function to the one obtained from the Gompertz distribution estimated above. \green{Since almost all the annuitants are older than $40$ at entry}, all the distributions are conditional on survival to age $40$. The survival functions are plotted as a function of age $x$ (for $x=40$ to $x=110$). The Gompertz curve is smooth whereas the KM is jagged. The figures clearly show that the estimated Gompertz model is a valid choice for approximating the KM curve.

\section{\zE{Dependence  Models}} \label{sec:copulas_definition}
\subsection{Background}
Copula models were introduced by Sklar \cite{sklar1959fonctions} in order to specify the  joint df of a random vector  by separating the behavior of the marginals and the dependence structure. Without loss of generality,  we focus on the bivariate case. We denote \zE{by}  $T (x_m)$ and $T(x_f)$  the future lifetime respectively for man and woman.  If $T (x_m)$ and $T(x_f)$  are positive and continuous, there exists a unique copula $C:\left[ 0,1\right]^2\rightarrow  \left[ 0,1\right]$ which specifies the joint df of the bivariate random vector $\left( T (x_m),T(x_f) \right) $ as follows
\BQNY 
			\mathbb{P}(T(x_m) \leqslant t_1, T(x_f)  \leqslant t_2)=C\left( \mathbb{P}\left( T(x_m) \leqslant t_1 \right),  \mathbb{P}\left( T(x_f) \leqslant t_2 \right)  \right) = C({_{t_1} q_{x_m}},{_{t_2} q_{x_f}}  ).
\EQNY
Similarly, the survival function of  $\left( T (x_m),T(x_f) \right) $ is written in terms of copulas and marginal survival functions. This is given by
\BQN \label{eq:SurvivalCopula}
			\mathbb{P}(T(x_m) > t_1, T(x_f)  >t_2)= \tilde{C}(   {_{t_1} p_{x_m}},{_{t_2} p_{x_f}}   )= {_{t_1} p_{x_m}}+ {_{t_2} p_{x_f}}  - 1 + C({_{t_1} q_{x_m}},{_{t_2} q_{x_f}}  ).
\EQN
A broad range of parametric copulas  has been developed in the literature. We refer to \green{ \cite{nelsen2007introduction} } for a review of the existing copula families. The Archimedean copula family is very popular in life insurance \green{ applications}, especially due to its flexibility in modelling dependent random lifetimes, \green{ see e.g., \cite{frees1996annuity,youn1999statistical} }.  If $\phi$ is  a convex and twice-differentiable strictly increasing function, the df of an Archimedean copula is given by
\BQNY 
			C_\phi(u,v)=\phi ^{-1}(\phi(u) + \phi(v)),
\EQNY
where $\phi: [0,1] \rightarrow [0,\infty]$ is the generator of the copula satisfying $\phi(1)=0$ with $ u,v \in [0,1]$. In this \red{paper}, \yT{four} well known copulas are discussed. Firstly, the Gumbel copula generated by
$$\phi(t)= (-\ln(t))^{-\alpha} , \quad  \alpha > 1,$$ 
which yields the copula
\BQN \label{eq:GumbelCopula}
C_\alpha(u,v)= \exp \{ -[ (-\ln(u))^{\alpha} + (-\ln(v))^{\alpha} ] ^{1/\alpha}  \}, \quad \alpha > 1.
\EQN
Secondly, we have the Frank copula 
\BQN \label{eq:FrankCopula}
		C_\alpha(u,v)= - \frac{1}{\alpha} \ln \biggl( 1+  \frac{(e^ {-\alpha u} - 1) (e^ {-\alpha v} - 1)  }{(e^ {-\alpha } - 1) } \biggr), \quad \alpha \neq 0,
\EQN
with generator 
$$\phi(t)=-\ln\biggl(    \frac{e^ {- \alpha t} - 1  }{e^ {- \alpha } - 1 } \biggr)  ,  \quad \alpha \neq 0.   $$ 
Thirdly, the Clayton copula is associated \zE{to} the generator 
$$\phi(t)= t^{-\alpha} -1, \quad  \alpha>0 $$ 
and is given by 
\BQN \label{eq:ClaytonCopula}
		C_\alpha(u,v)= (u^{-\alpha} + v^{-\alpha} -1 )^{-1/ \alpha}, \quad \alpha >0.
\EQN
Finally, the Joe copula 
\BQN \label{eq:JoeCopula}
C_\alpha(u,v)= 1-\Bigl((1-u)^\alpha + (1-v)^\alpha -(1-u)^\alpha  (1-v)^\alpha \Bigr)   ^{1/\alpha}  , \quad \alpha > 1
\EQN
\eH{has generator $\phi(t)= -\ln(1-(1-t)^{-\alpha}) , \alpha > 1.$}

\zE{Clearly, the} parameter $\alpha$ in  \eqref{eq:GumbelCopula}\zE{-}\eqref{eq:JoeCopula} determines the dependence level  between the two marginal distributions. In our case, that would be the lifetimes of wife and husband.  Youn and Shemyakin \cite{youn1999statistical} have utilized a Gumbel copula where the association parameter $\alpha$ depends on $d$ as  follows
\begin{equation}
\label{eq:paramYoun}
\alpha(d) = 1+  \frac{\beta_0}{1 +  \beta_2 d^2 }, \,\;\;\; \beta_0, \beta_2 \in \R
\end{equation}
where $d=x_m -x_f$ with $x_m$ and $x_f$ the ages for male and female, respectively. \\
\zE{In our model for $\alpha$}, in addition to this specification, the \eH{gender} of the elder partner, represented by the sign of $d$, is also taken into account. This latter is captured through the second term of the denominator $\beta_1 d$ in equations \eqref{eq:paramFrankClayton} and \eqref{eq:paramGumbel}. Thus, \zE{for our model} the copula association parameter for the Frank and the Clayton is expressed by
			\BQN \label{eq:paramFrankClayton}
					\alpha(d) =  \frac{\beta_0}{1 + \beta_1 d + \beta_2 \abs{d} }, \quad \beta_0,\beta_1, \beta_2 \in \R.
			\EQN	
		Since the copula parameter $\alpha$ in the Gumbel and  Joe copulas is restricted to be greater than $1$, the corresponding dependence parameter in \eqref{eq:paramGumbel} is allowed to have an intercept of $1$ and we write
			\BQN \label{eq:paramGumbel}
					\alpha(d) = 1+  \frac{\beta_0}{1 + \beta_1 d + \beta_2 \abs{d} }, \quad \beta_0,\beta_1, \beta_2 \in \R.
			\EQN
It can be seen that if $\beta_1 < 0$, the dependence parameter is lower when husband is younger than wife, i.e. $d<0$. Also when $d$ tends to infinity, the dependence parameter goes to $0$ for Frank and Clayton and $1$ for the Gumbel copula, thus tending towards the independence assumption. \zE{Note in passing that instead of taking} $d^2$ as in equation \eqref{eq:paramYoun}, we use $ \abs{d}$ in both \eqref{eq:paramFrankClayton} and \eqref{eq:paramGumbel} for the representation of the absolute age difference.

\subsection{\zE{Estimation of Parameters }}	
The maximum likelihood procedure has been widely used to fit lifetime data to copula models, see e.g., \green{  \cite{lawless2011statistical, shih1995inferences,carriere2000bivariate}}. A priori, this method consists in estimating jointly the marginal and copula parameters at once. However, given the \zE{huge} number of parameters to be estimated at the same time, \zE{this approach is} computationally intensive.  \zE{Therefore},  we adopt a procedure that allows the determination of marginal and copula parameters, separately. In this respect, Joe and Xu \cite{joe1996estimation} have proposed a two step technique which, firstly estimates the marginal parameters $\theta_j, j = m,f,$  and the copula parameter $\alpha(d)$ in the  second step.
\COM{
A priori, le lecteur serait tenter de croire que on devrait tout estimerd jointly d'un coup; et écrire la formule de la likelihood dans ce cas. mais ce sera computationnellement intensive.
Pourquoi dur computationnellement? Stabilité des paramètres, requiert au moins 7 paramètres à estimer. et la stabilité des paramètres laisse à désirer. Une solution est le modèle de Jo qui propose de faire l'estimation en deux étapes. D'abord estimer séparément les marginals et espliquer clairement en faisant référence à la première partie. et deuxièment, le copula. Ce qui réduit notre modèle en 3 modèles: les marginals avec deux paramètres chacun et le joint avec 3 paramètres. En plus de cela, ajouter un commentaire en terme de fiabilité des résultats. préciser aussi que le omnibus ici sert à vérifier nos résultats. Expliquer clairement la particularité du Omnibus en faisant allusion à l'IFM
} 
This is referred to as the \textit{inference functions for margins} (IFM) method. \zE{Specifically,}  the  survival function of each lifetime is evaluated by maximazing the likelihood function in \eqref{eq:marginal_likelihood}. For each couple $i$ with $x_m^i$ and $x_f^i$, let $u_i:= {_{t_m^{i}}}p_{x_m^i} ( \hat{\theta}_m) $ and $v_i:= {_{t_f^{i}}}p_{x_f^i} ( \hat{\theta}_f) $  be the resulting marginal survival functions for male and female, respectively. Considering the right-censoring feature of the two lifetimes as indicated by $\delta_m^i$ and $\delta_f^i$, the estimates $\widehat { \alpha(d)}$ of the copula parameters are obtained by maximizing  the  likelihood function 
\BQN \label{eq: likelihoodCopulaIFM}
		L( \alpha(d)):= 	L( \alpha)& =&
		 \prod_{i=1}^{n}\;\;\;
		 \Biggr[
		\frac{\partial ^2 \tilde{C}_{\alpha} (u_i,v_i )   }{\partial u_i \partial v_i } \Biggl] ^{\left( 1-\delta_m^i\right)  \left( 1-\delta_f^i\right) }\;\;\;
		\Biggr[\frac{\partial \tilde{C}_{\alpha}(u_i,v_i ) }{\partial u_i)  }  \Biggl] ^{\left( 1-\delta_m^i\right)\delta_f^i} 
		 \notag \\
		 && \;\;\;\times \;\;
		  \Biggr[\frac{\partial \tilde{C}_{\alpha}(u_i,v_i )  }{ \partial v_i }  \Biggl] ^{\delta_m^i \left( 1-\delta_f^i\right)} \;\;\;\;\;\;
		  \Biggr[ \tilde{C}_{\alpha}(u_i,v_i ) \Biggl] ^{\delta_m^i \delta_f^i }.
\EQN
A similar two-step technique, known as the \textit{Omnibus semi-parametric  procedure} or the \textit{pseudo-maximum likelihood}, was also introduced by Oakes \cite{oakes1989bivariate}. In this procedure, the marginal distributions are considered as nuisance parameters of the copula model. The first step consists in estimating the two marginals survival functions non-parametrically using the KM method. After rescaling the resulting estimates by $\frac{n}{n+1}$, we obtain the pseudo-observations $ ( U_{i,n} , V_{i,n})$ where
\begin{equation}
U_{i,n} = \frac{\hat{S}_m( x^i_m +t^i_{m} ) }{\hat{S}_m( x^i_m) } \;\;\;\;\; \text{   and   }\;\;\;\;\;  V_{i,n} = \frac{\hat{S}_m( x^i_f +t^i_{f} ) }{\hat{S}_m( x^i_f ) }.
\nonumber
\end{equation}
In the second step, the copula estimation is achieved by maximizing the following function 
 \BQN \label{eq: likelihoodCopulaPseudo}
		L( \alpha(d)):= 		L( \alpha)& =&
		 \prod_{i=1}^{n} \;\;\;
		 \Biggr[
		\frac{\partial ^2 \tilde{C}_{\alpha} (U_{i,n},V_{i,n} )   }{\partial U_{i,n} \partial V_{i,n} } \Biggl] ^{\left( 1-\delta_m^i\right)  \left( 1-\delta_f^i\right) }\;\;\;
		\Biggr[\frac{\partial \tilde{C}_{\alpha}(U_{i,n},V_{i,n}) }{\partial U_{i,n}  }  \Biggl] ^{\left( 1-\delta_m^i\right)\delta_f^i}
		 \notag \\
		 && \;\;\; \times \;\;
		  \Biggr[\frac{\partial \tilde{C}_{\alpha}(U_{i,n},V_{i,n} )  }{ \partial V_{i,n} }  \Biggl] ^{\delta_m^i \left( 1-\delta_f^i\right)} \;\;\;\;\;\; 
		  \Biggr[ \tilde{C}_{\alpha}(U_{i,n},V_{i,n} ) \Biggl] ^{\delta_m^i \delta_f^i }.
\EQN
Genest et al. \cite{genest1995semiparametric} and Shih and Louis \cite{shih1995inferences} have shown that the stemmed estimators of the copula parameters  are consistent and asymptotically normally distributed. Due to their computational advantages, the IFM and the Omnibus approaches are used in our estimations. By comparing the results stemming from the two techniques, we can analyze to which extent a certain copula is a reliable model for bivariate lifetimes within a couple. Table \ref{table: IFM estimates copula params} and Table \ref{table: omnibus estimates copula params} display the copula estimations based on our dataset. 
The estimated values from the IFM and the omnibus estimations are quite close for the Gumbel, the Frank and the Joe copulas. The important difference observed in the Clayton case indicates that \green{ this copula} is probably not appropriate for modelling the bivariate lifetimes in our dataset. The negative sign of $\hat{\beta}_1$ in all cases demonstrates that if husband is older than wife (i.e. $d>0$), their lifetimes are more likely to be correlated. The positive sign of $\hat{\beta}_2$ suggests that the higher the age difference is, the lesser is the level of dependence between lifetimes. The parameters $\hat{\beta}_1$ and $\hat{\beta}_2$ have opposing \green{ effects} on $\hat{\alpha}\left( d\right) $. That is why the maximum level of dependence is attained when $d=0$, i.e. when wife and husband have exactly the same age. 
 \COM{
 	\begin{center}	
		\begin{tabular}{|c||c|c|c|| c| c|c|c|| }
			\hline
			& \multicolumn{3}{c||}{IFM} & \multicolumn{3}{c| }{Omnibus} \\
		\hline
		
	Copula 	&  $ \hat{\beta}_0$  & $\hat{\beta}_1$  & $\hat{\beta}_2$  &  $\hat{\beta}_0$  & $\hat{\beta}_1$  & $\hat{\beta}_2$   \\

		\hline
	Gumbel & 1.027 & -0.024 &  0.0361 &0.976 &-0.022 &  0.030 \\
	\hline 
	Frank  &  7.359 & -0.017 & 0.023  & 7.294  & -0.016 & 0.021    \\
	\hline 
	Clayton  &  2.461 & -0.302 & 0.464  & 1.924  & -0.169 & 0.296    \\
	\hline 
		\end{tabular}
		\captionof{table}{ Copulas parameters estimate of the dependence parameters in $\alpha(d)$.} \label{table:estimates copula params}
\end{center}
}

\begin{center}	
		\begin{tabular}{|c||c|c|c|| c| c|c||c| }
	\hline
	\multirow{2}{*}
	{Copula parameters}		& \multicolumn{6}{c||}{$\alpha(d)$} & $\alpha$ \\
	\cline{2- 8} 	&  $ \hat{\beta}_0$  & $\hat{\beta}_1$  & $\hat{\beta}_2$  &  $\hat{\alpha}(-2)$ & $\hat{\alpha}(0)$& $\hat{\alpha}(2)$  & $\hat{\alpha}$   \\
	\hline
	Gumbel & 1.027 & -0.024 &  0.036  & 1.917  & 2.027 & 2.003 & 1.993  \\
	\hline 
	Frank  &  7.359 & -0.017 & 0.023  	  & 6.813  &  7.359  &  7.272 & 7.065    \\
	\hline 
	Clayton  &  2.461 & -0.302 & 0.464  & 0.972 & 2.461 &  1.857 & 1.960    \\
	\hline 
	Joe & 1.488  & -0.063 &  0.063  & 2.189  & 2.488  &  2.488  &  2.389    \\
	\hline 	
	\end{tabular}
	\captionof{table}{ IFM method: copula parameters estimate  $\alpha(d)$ and $\alpha$.} \label{table: IFM estimates copula params}
\end{center}
\begin{center}	
		\begin{tabular}{|c||c|c|c|| c| c|c||c| }
	\hline
	\multirow{2}{*}
	{Copula parameters}		& \multicolumn{6}{c||}{$\alpha(d)$} & $\alpha$ \\
	\cline{2- 8} 	&  $ \hat{\beta}_0$  & $\hat{\beta}_1$  & $\hat{\beta}_2$  &  $\hat{\alpha}(-2)$ & $\hat{\alpha}(0)$& $\hat{\alpha}(2)$  & $\hat{\alpha}$   \\
	\hline
	Gumbel & 0.976 & -0.022 &  0.030  & 1.884  & 1.976 & 1.960 & 1.924  \\
	\hline 
	Frank  &  7.294 & -0.016 & 0.021  	  & 6.791  &  7.294  &  7.223 & 6.828    \\
	\hline 
	Clayton  &  1.924 & -0.169 & 0.296  &  0.997 & 1.924 &  1.534 & 1.117    \\
	\hline 
	Joe & 1.409 & -0.0505 & 0.0581  &  2.158   & 2.409 &  2.388 & 2.352  \\
	\hline 
	\end{tabular}
	\captionof{table}{ Omnibus approach: copula parameters estimate  $\alpha(d)$ and $\alpha$.} \label{table: omnibus estimates copula params}
\end{center}

Our estimate of $\alpha(d)$ under the Gumbel copula is quite similar to the results in the model of Youn and Shemyakin  \cite{youn1999statistical} where $\hat{\beta}_0 = 1.018, \hat{\beta}_1 = 0 $ and $\hat{\beta}_2 = 0.021 $. Column $8$ contains the estimation output when the dependence parameter $\alpha$ does not depend on $d$. When $d=0$, $\alpha\left( 0\right) =\beta_0$ (or $1+\beta_0$ for Gumbel \yT{and Joe}) and that is equivalent to the case where the dependence parameter is not in function of the age difference. By comparing the sixth and the eighth columns, it can be seen that the model without age difference underestimates the lifetime dependence level between spouses.

\subsection{Goodness of fit}
A goodness of fit procedure is performed in order to assess the robustness of our model. For this purpose, the model, including age difference and \eH{gender} of the elder member within the couple with $\alpha\left( d\right) $, is compared to two other types, namely the one where the copula parameter does not depend on $d$ and the model of Youn and Shemyakin \cite{youn1999statistical}. Many approaches  for testing the goodness of fit of copula models are proposed in the litterature, see e.g.,  \cite{genest2009goodness, berg2009copula}. We refer to  \cite{genest2009goodness} for an  overview of the existing methods. 
\COM{In our framework, the goodness of fit approach is based on the non parametric empirical copula process $C_n^o$ given by
			\BQN \label{eq:original empirical copula}
		 		C_n^o(u_1,u_2) = 
		 		\frac{1}{ n }  \sum_{i=1}^n 
		 		\mathbbm{1}_ { \{T^i_{m}  \leqslant \hat{F}_{m,n} ^ {-1} (u_1) , T^i_{f}  \leqslant \hat{F}_{f,n} ^{-1} (u_2)   \} }, \;\;\;\;\; u_1, u_2 \in \left[ 0,1\right] \nonumber
		 	\EQN
where $\hat{F}_{j,n}^{-1}$ is the KM estimator of the quantile function of the observation $T_{j}^i, j=m,f $. }There are several contributions highlighting the properties of the empirical copula, especially when the data are right censored, the contributions  \cite{dabrowska1988kaplan, prentice2004hazard,gribkova2015non} are some examples. In our framework, the goodness of fit approach is based on the non parametric  copula introduced by Gribkova et al. \cite{gribkova2015non} 	as follows
		 	\BQN \label{eq:empirical copula}
		 		C_n(u_1,u_2) = 
		 		\frac{1}{ n }  \sum_{i=1}^n 
		 		(1-\delta_{m}^i) (1-\delta_{f}^i)  W_{in} 
		 		\mathbbm{1}_ { \{T{(x_m^i)}  \leqslant \hat{F}_{m,n} ^ {-1} (u_1) , T{(x_f^i) }  \leqslant \hat{F}_{f,n} ^{-1} (u_2)   \} }, 
		 	\EQN
where
\green{ $W_{in} = \frac{1}{ S_{B_m} (\max(T_{m}^i, T_{f}^i-\epsilon_i ) -) }$ and $S_{B_m}$ is the survival function of the right censored random variable $B_m$ } that is estimated using KM approach; $\epsilon_i =B_f^i - B_m^i$.  The term $\hat{F}_{j,n}^{-1}$ is the KM estimator of the quantile function \green{of}  $T(x_j^i), j=m,f $.
The particularity of equation \eqref{eq:empirical copula} is that, the uncensored observations are twice weighted (with  $1/n$ and $W_{in}$) unlikely to the original empirical copula where the same weight $1/n$ is assigned to each observation. The weight $W_{in}$ is devoted to compensate right censoring. Based on the p-value, the goodness of fit test indicates to which extent a certain parametric copula is close to the empirical copula $C_n$. We adopt the Cram\`er-von Mises statistics to assess the adequacy of the hypothetical copula to the empirical one, \zE{namely} 
 			 	\BQN \label{eq: von_mises_def} 
		 			 				\zE{\mathcal{V}_n}= \int_{[0,1]^2}  \zE{K_n}(v) dK_n(v),
		 		\EQN
		 			 	where 	 $K_n(v)=\sqrt{n} (C_n(v) - C_{\hat{\alpha}(d)}(v) )$	is the empirical copula process. Genest et al. \cite{genest2009goodness} have proposed an empirical version of equation \eqref{eq: von_mises_def} which is given by
		 	\BQN \label{eq:Von mises Empirical }
		 		\widehat{\mathcal{V}}_n= \sum_{i=1}^n ( C_n(u_{1i},u_{2i}) - C_{\hat{\alpha}(d)}(u_{1i},u_{2i}))^2.	 	
		 	\EQN
The assertion, the bivariate lifetime within the couple is described by the studied copula, is then tested under the null hypothesis $H_0$. Since the Cram\`er-von Mises statistics \red{$\widehat{\mathcal{V}}_n$}
 does not possess an explicit df, we implement a bootstrap procedure to evaluate the p-value as presented in the following pseudo-algorithm. For some large integer $K$, the following \green{steps are}  repeated for every $k=1,\ldots,K$:
		 		\BIT
		 	\item \textbf{Step 1} 
		 	Generate lifetimes from the hypothetical copula, i.e. $(U_i^b, V_i^b), i=1,\ldots,n$ is generated from  $C_{\hat{\alpha}(d) }$. If the IFM method is used to determine $\hat{\alpha}(d)$, then the two lifetimes are produced from the Gompertz distribution  
		 		$$ ( t^{b,i}_m = F^{-1}_{x_m}(U_i^b,\hat{\theta}_m), t^{b,i}_f = F^{-1}_{x_f}(V_i^b,\hat{\theta}_f)),$$ 
		 	   where  $\hat{\theta}_j,\; j=m,f$ are taken from Table \ref{table: Gompertz_estimates},
		 	   while, for the omnibus, the corresponding lifetimes are generated with the KM estimators of the quantile functions  of $T\left( x_j\right),\; j=m,f $
		 	  $$ ( t^{b,i}_m = \hat{F}_{m,n} ^ {-1}(U_i^b),\; t^{b,i}_f = \hat{F}_{f,n} ^ {-1}(V_i^b)).$$ 
		 	\item \textbf{Step 2} 
		 	Generate the censored variables  \green{$B_{m}^{b,i}$ and $B_{f}^{b,i}, i=1,\ldots,n$} from the empirical distribution of $B_{m}$ and $ B_{f}$ respectively.
		 	\item \textbf{Step 3} 
		 	Considering the same data as used for the estimation, replicate the insurance portfolio by calculating 
		 	$$T^b(x_m^i)= \min(t^{b,i}_m,   \green{B_{m}^{b,i}} ), \quad \delta^{b,i}_m=\mathbbm{1}_{\{t^{b,i}_m \geqslant \green{B_{m}^{b,i}} \} } ,$$
		 	$$T^b(x_f^i)= \min(t^{b,i}_f, \green{B_{f}^{b,i}} ), \quad \delta^{b,i}_f=\mathbbm{1}_{\{t^{b,i}_f \geqslant \green{B_{f}^{b,i}}  \} } $$
		 	for each couple $i$ of ages $x_m^i$ and $x_f^i$.
		 		\item \textbf{Step 4}
		 		If  the IFM approach is chosen in $\textbf{Step 1}$, the parameters of the marginals and  the hypothetical copula parameters are estimated  from the bootstrapped data $(T^b(x_m^i),T^b(x_f^i), \delta^{b,i}_m, \delta^{b,i}_f)$ by maximizing \eqref{individual_likelihood} and  \eqref{eq: likelihoodCopulaIFM} whereas under the omnibus approach, the hypothetical copula parameters are estimated  from the bootstrapped data as well by maximizing  equation \eqref{eq: likelihoodCopulaPseudo}.
		 	 \item \textbf{Step 5}
		 	 Compute the Cram\`er-von Mises statistics $\widehat{\mathcal{V}}^b_{n,k}$ using \eqref{eq:Von mises Empirical }.
		 	 \item \textbf{Step 6}
		 Evaluate the estimate of the p-value as follows
		 	$$\hat{p} = \frac{1}{K+1} \sum_{k=1}^ {K} \mathbbm{1}_ { \{\widehat{\mathcal{V}}_{n,k}^b \geqslant \widehat{\mathcal{V}}_n \}}.$$
		 	 	\EIT

Based on $1000$ bootstrap samples\yT{,} the results of the goodness of fit is summarized in Table \ref{table:GOF results}. It can be seen that for both IFM and Omnibus, our model have a greater p-value than the model without age difference, showing that age difference between spouses is an important dependence factor of their joint lifetime. Under the Gumbel model in Youn and Shemyakin \cite{youn1999statistical} where $\beta_1=0$, the p-value is evaluated at $0.678$. For the Gumbel copula in Table \ref{table:GOF results}, the p-value in the model with $\alpha\left( d\right) $ is slightly higher, strenghthening the evidence that the sign of $d$ captures some additional association between spouses.
\begin{center}	
		\begin{tabular}{|c||c|c||c| c|  }
			\hline
			& \multicolumn{2}{c||}{IFM} & \multicolumn{2}{c| }{Omnibus} \\
		\hline		
	Copula parameters	&  $\alpha$ & $\alpha(d)$  	&  $\alpha$ & $\alpha(d)$ \\
		\hline
	Gumbel & 0.647 & 0.679 & 0.639   &  0.670 \\
	\hline 
	Frank  &  0.518 & 0.525 &  0.521  & 0.530     \\
	\hline 
	Clayton  &  0.111 & 0.163 & 0.120   & 0.158     \\
	\hline 
	Joe  & 0.321  & 0.338  &  0.318   &  0.329   \\
	\hline 
		\end{tabular}
		\captionof{table}{ Goodness of fit test: p-value of each copula model.} \label{table:GOF results}
\end{center}

At a critical level of $5 \%$, the three copula families are accepted, even though the  Clayton copula performs inadequately. 
Actually, as pointed out in \green{ \cite{gribkova2015non}}, the important percentage of censored data in the sample results in a huge loss of any GoF test. Therefore, these results can not efficiently assess the lifetime dependence within a couple. Nevertheless, the calculated p-values may give an idea about which direction to go. \red {In this regards}, since the Gumbel and Frank copulas have the highest p-value, they are good candidates for addressing the dependence of the future lifetimes of husband and wife in this Canadian life insurer portfolio.
\section{Insurance applications}
\subsection{Joint life insurance contracts}\label{subSec:defProb}		
Multiple life actuarial calculations is common in the insurance practice. Hereafter, $\left( x\right) $ stands for the husband aged $x$ whereas $\left( y\right) $ is the wife. Considering a couple $\left( xy\right) $, $T\left( xy\right) $ describes the remaining time until the first death between $\left( x\right)$ and $\left( y\right)$ and, it is known as the \textit{joint-life status}. Conversely, $T\left(\overline{ xy}\right)$ is the time until death of the \textit{last survivor}. The variables $T\left(\overline{ xy}\right)$ and $T\left( xy\right)$ are random and we can write
\begin{equation}
T\left( xy\right)= \min \left( T\left( x\right) , T\left( y\right) \right) \text{   whereas   }T\left(\overline{ xy}\right)= \max \left( T\left( x\right) , T\left( y\right) \right).
\nonumber
\end{equation}
As in the single life model, the survival probabilities are given by
\begin{equation}
_tp_{xy}=\mathbb{P}\left( T\left( xy\right)  > t \right) \;\;\;\;\; \text{     and    } \;\;\;\;\; _tp_{\overline{xy}}=\mathbb{P}\left( T\left( \overline{xy}\right)  > t \right).\end{equation}
\eH{Clearly, if} $T\left( x\right)$ and $ T\left( y\right)$ are independent, then 
\begin{equation}
{_t}p_{xy}= {_t}p_{x} \; {_t}p_{y} \;\;\;\;\; \text{     and    } \;\;\;\;\; {_t}p_{\overline{ xy }}=1 -  {_t}q_{x} \; {_t}q_{y}.
\nonumber
\end{equation}
The curtate life expectancies, for $T\left( xy\right) $ and $T\left( \overline{xy}\right) $ respectively, are given by
\begin{equation}
\mathit{e}_{xy}=\mathbb{E}\left( T\left( xy\right) \right) =\sum_{t=1}^{\infty} {_t}p_{xy}     \;\;\;\;\; \text{     and     }      \;\;\;\;\;        \mathit{e}_{\overline{xy}}=\mathbb{E}\left( T\left( \overline{xy} \right) \right) =\sum_{t=1}^{\infty} {_t}p_{\overline{xy}},
\nonumber
\end{equation}
with the following relationship
\begin{equation}
\mathit{e}_{\overline{xy}} = \mathit{e}_{x} + \mathit{e}_{y} - \mathit{e}_{xy}.
\nonumber
\end{equation}
Figures \ref{life_expectancy_comparison} and \ref{life_expectancy_comparison_1} compare the evolution of $\mathit{e}_{\overline{xy}}$ as a function of the age difference $d=x-y$, under the following models:
\begin{itemize}
\item Model A: $T\left( x\right) $ and $T\left( y\right) $ are independent;
\item Model B: $T\left( x\right) $ and $T\left( y\right) $ are dependent with a constant copula parameter $\alpha=\alpha_0$;
\item Model C: $T\left( x\right) $ and $T\left( y\right) $ are dependent with a copula parameter $\alpha\left( d\right) $ as described in \eqref{eq:paramFrankClayton} and  \eqref{eq:paramGumbel}.

\end{itemize}

On the left (resp. right), the graphs were constructed under the assumption of $x=65$ (resp. $y=65$) for the husband (resp. wife) and the age difference $d$ ranges from $-20$ to $20$ as more than $99\%$ of our portfolio belongs to this interval. The fixed age is set to $65$ because this is the retirement age in many countries. The analysis was made under the four families of copula described in Section \ref{sec:copulas_definition}. In general, it can be seen that the life expectancy of the last survivor $\mathit{e}_{\overline{xy}}$ increases when $\mathit{e}_{\overline{xy}}=\mathit{e}_{\overline{65:65-d}}$ whereas it decreases when $\mathit{e}_{\overline{xy}}=\mathit{e}_{\overline{65+d:65}}$. This result strengthens the evidence that the sign of $d$ has an effect on \green{annuity values}. For example, when  $\abs{d}=10$ under the Gumbel copula,
\begin{equation}
\mathit{e}_{\overline{65:55}}=32.62 \geq \mathit{e}_{\overline{55:65}}=28.82.
\nonumber
\end{equation} 

\begin{figure}[H]\label{life_expectancy_comparison}

\begin{minipage}[c]{.46\linewidth}
\includegraphics[width=7.5cm,height=7.5 cm]{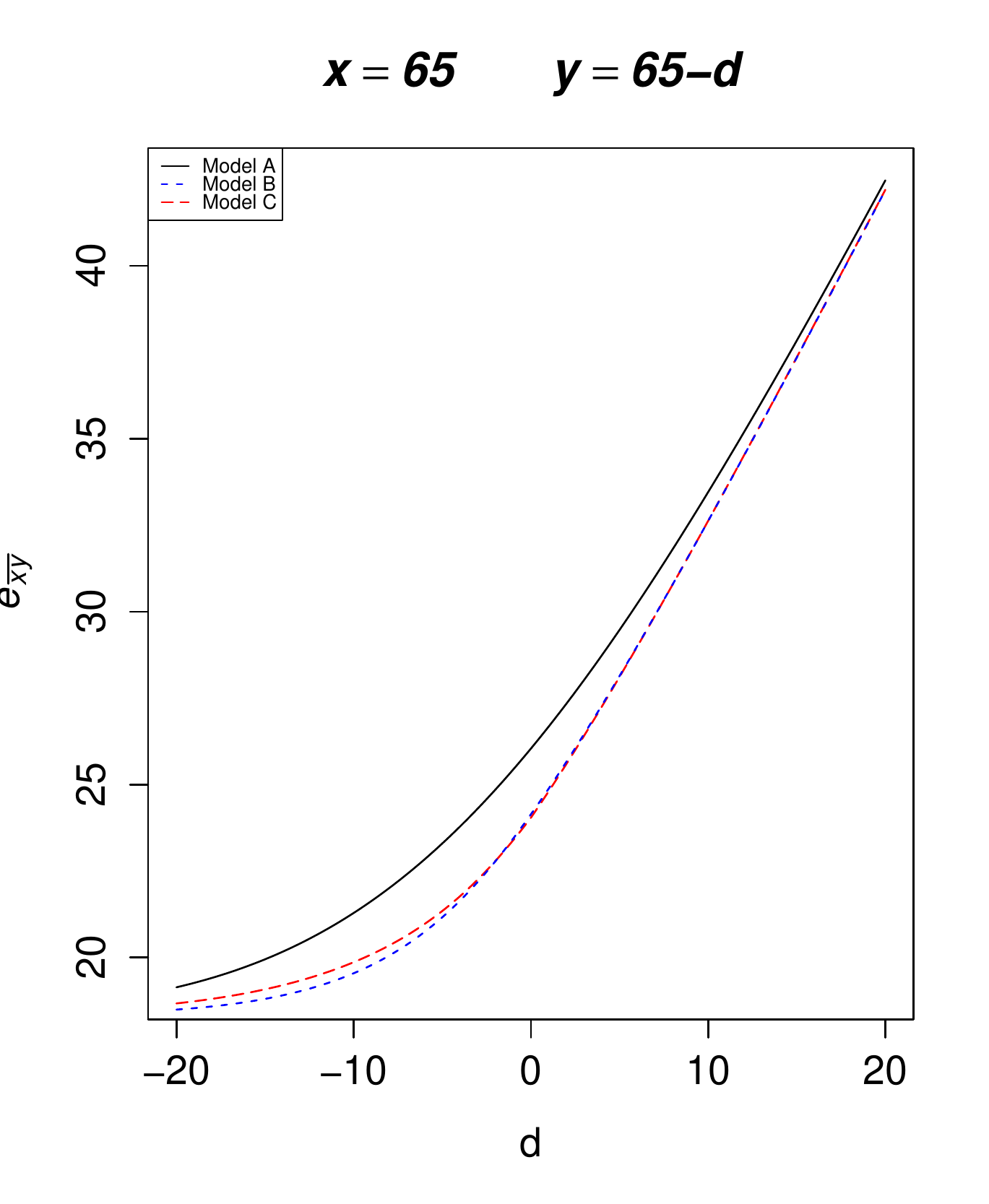}
\vspace{-0.75cm}
\label{ageddpdceratio1_10scenarios_sanstitre}
\subcaption{Gumbel copula: $x=65$}
\end{minipage} \hfill
\vspace{-0.35cm}
\begin{minipage}[c]{.46\linewidth}
\includegraphics[width=7.5cm,height=7.5 cm]{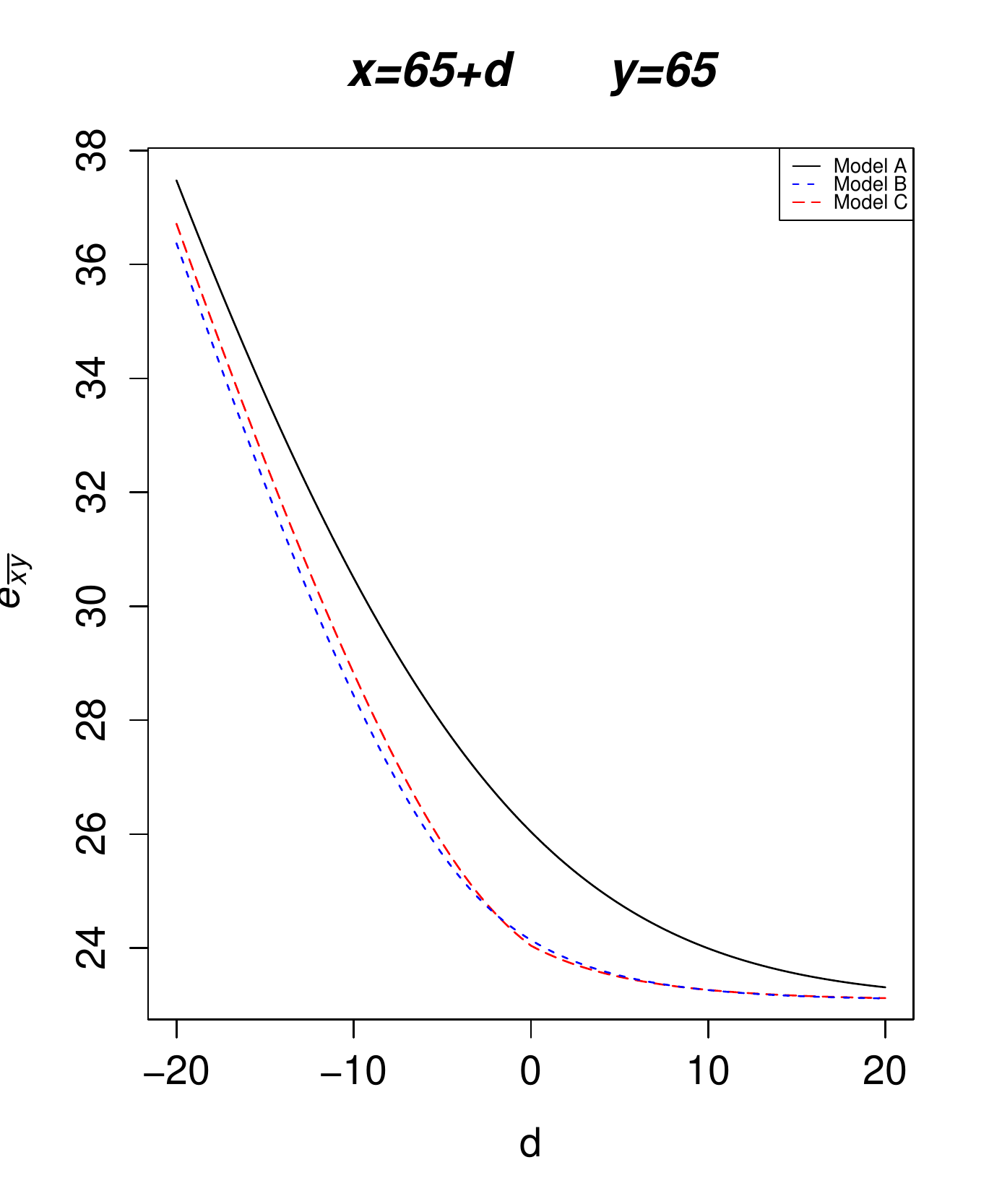}
\vspace{-0.75cm}
\label{ageddpdceratio1_10scenarios_sanstitre}
\subcaption{Gumbel copula: $y=65$}
\end{minipage}
\vspace{0.25cm}

\begin{minipage}[c]{.46\linewidth}
\includegraphics[width=7.5cm,height=7.5 cm]{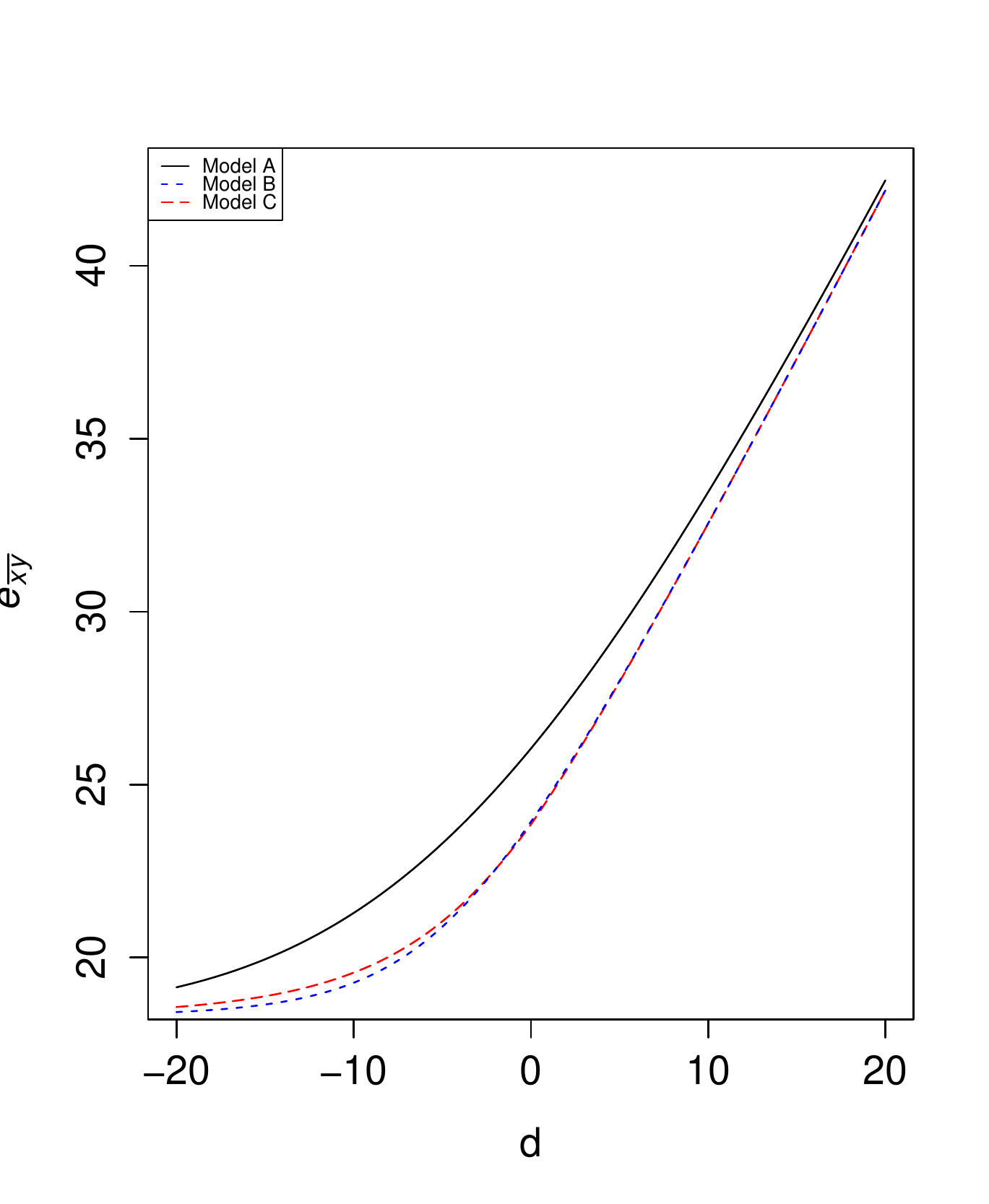}
\vspace{-0.75cm}
\label{ageddpdceratio1_10scenarios_sanstitre}
\subcaption{Frank copula: $x=65$}
\end{minipage} \hfill
\vspace{-0.35cm}
\begin{minipage}[c]{.46\linewidth}
\includegraphics[width=7.5cm,height=7.5 cm]{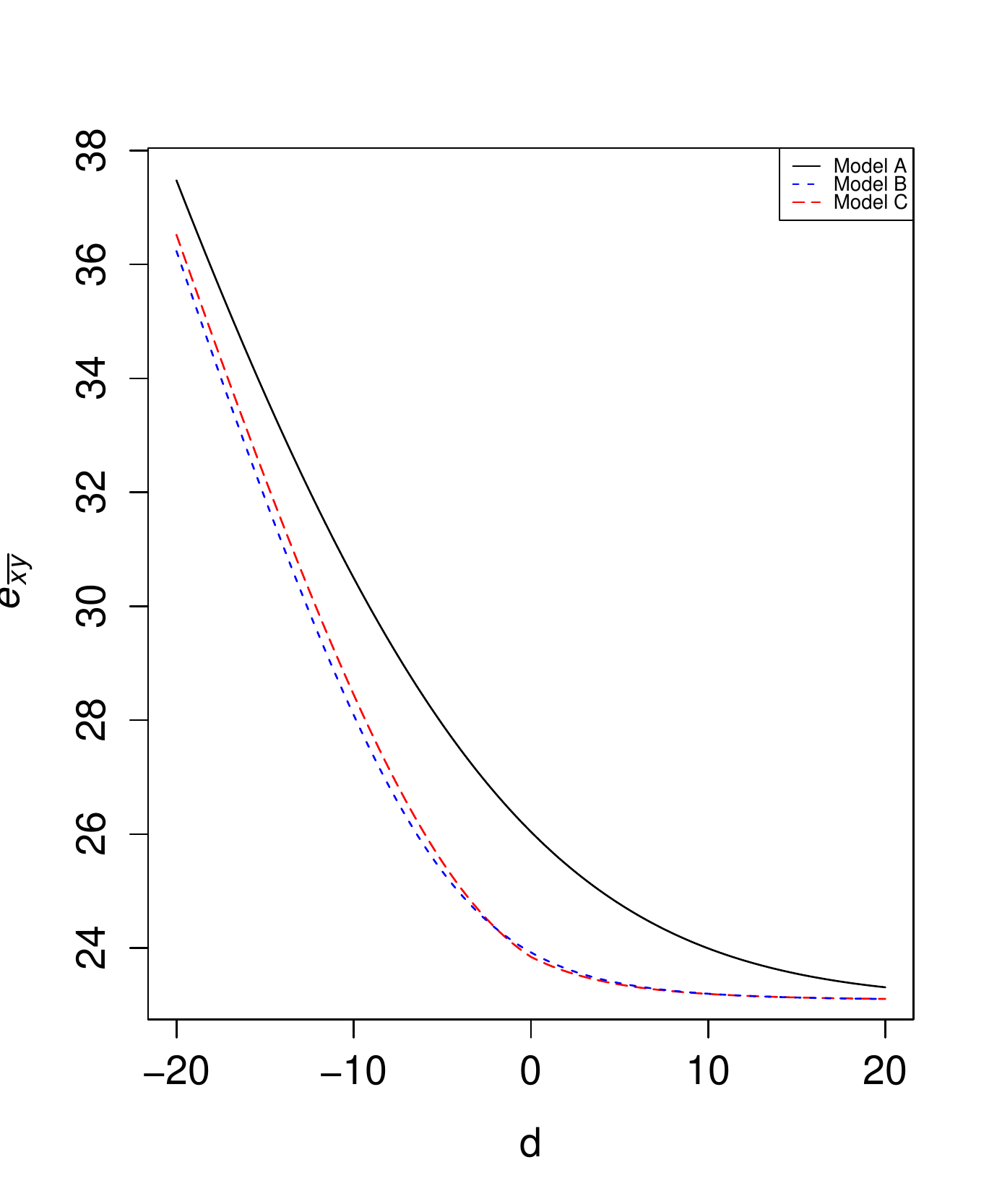}
\vspace{-0.75cm}
\label{ageddpdceratio1_10scenarios_sanstitre}
\subcaption{Frank copula: $y=65$}
\end{minipage}
\vspace{0.25cm}

\caption{Comparison of $\mathit{e}_{\overline{xy}}$ under model A, B and C: Gumbel and Frank copulas}
\label{life_expectancy_comparison}
\end{figure}

\begin{figure}[H]\label{life_expectancy_comparison_1}

\begin{minipage}[c]{.46\linewidth}
\includegraphics[width=7.5cm,height=7.5 cm]{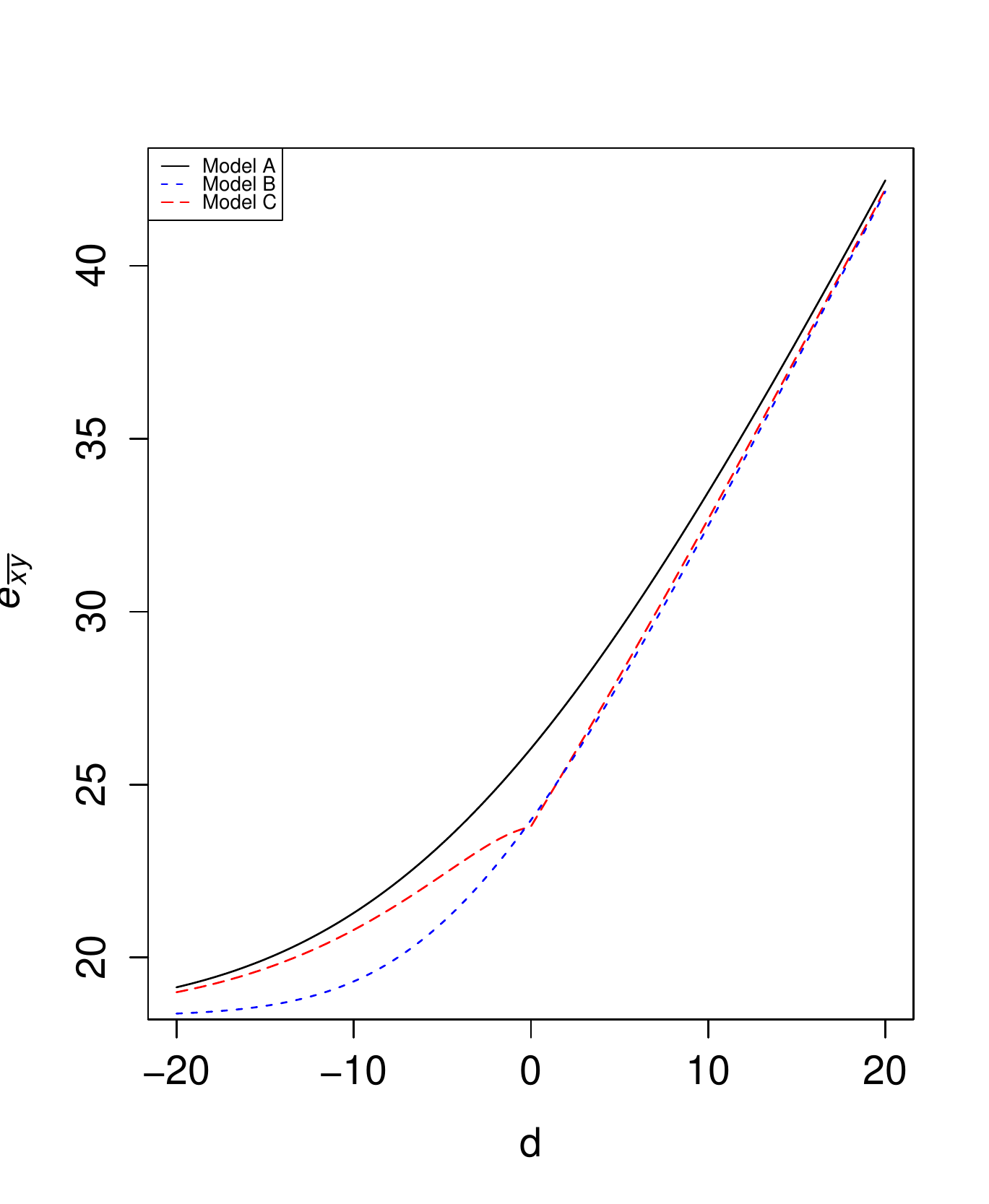}
\vspace{-0.75cm}
\label{ageddpdceratio1_10scenarios_sanstitre}
\subcaption{Clayton copula: $x=65$}
\end{minipage} \hfill
\vspace{-0.35cm}
\begin{minipage}[c]{.46\linewidth}
\includegraphics[width=7.5cm,height=7.5 cm]{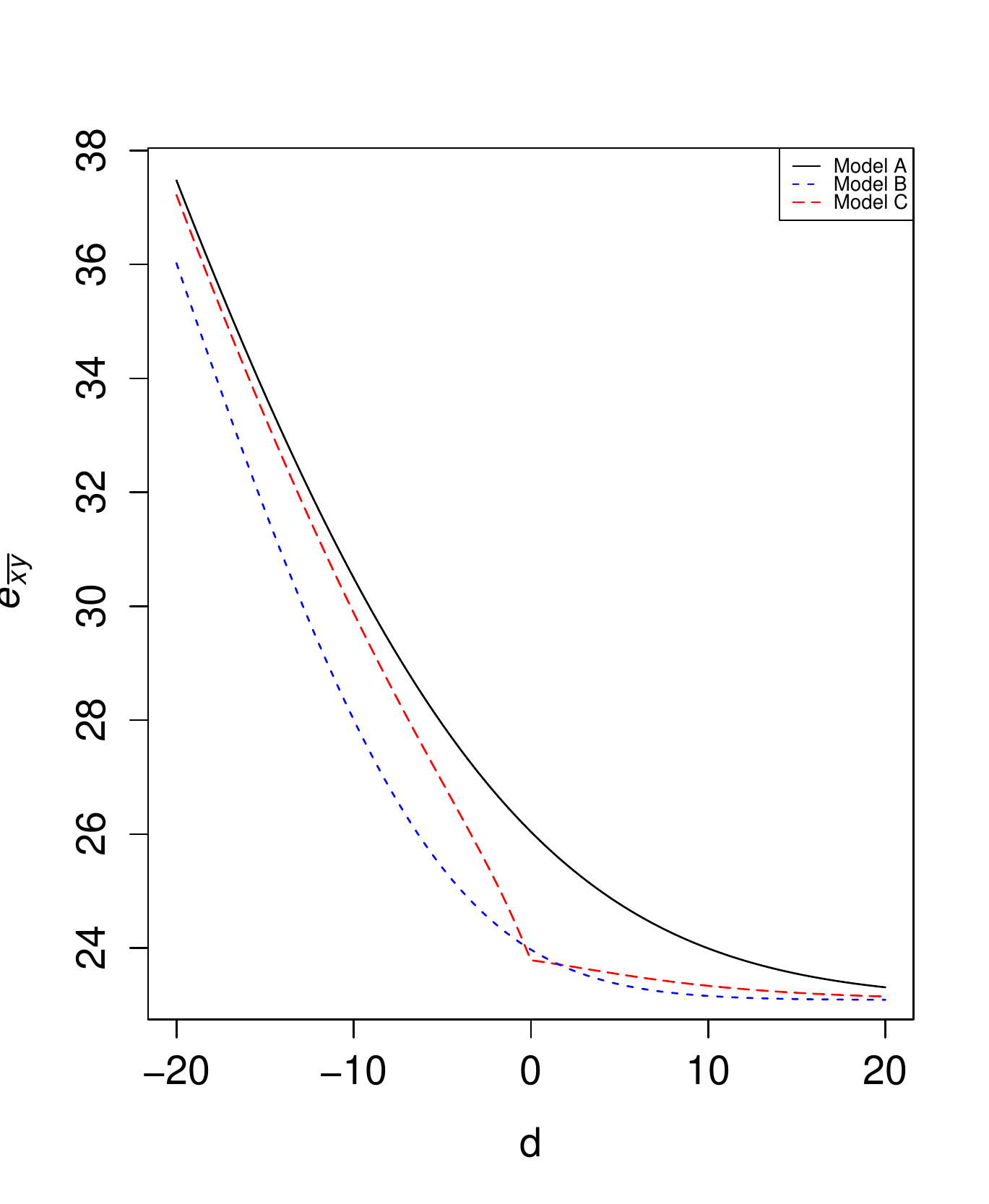}
\vspace{-0.75cm}
\label{ageddpdceratio1_10scenarios_sanstitre}
\subcaption{Clayton copula: $y=65$}
\end{minipage}
\vspace{0.25cm}

\begin{minipage}[c]{.46\linewidth}
\includegraphics[width=7.5cm,height=7.5 cm]{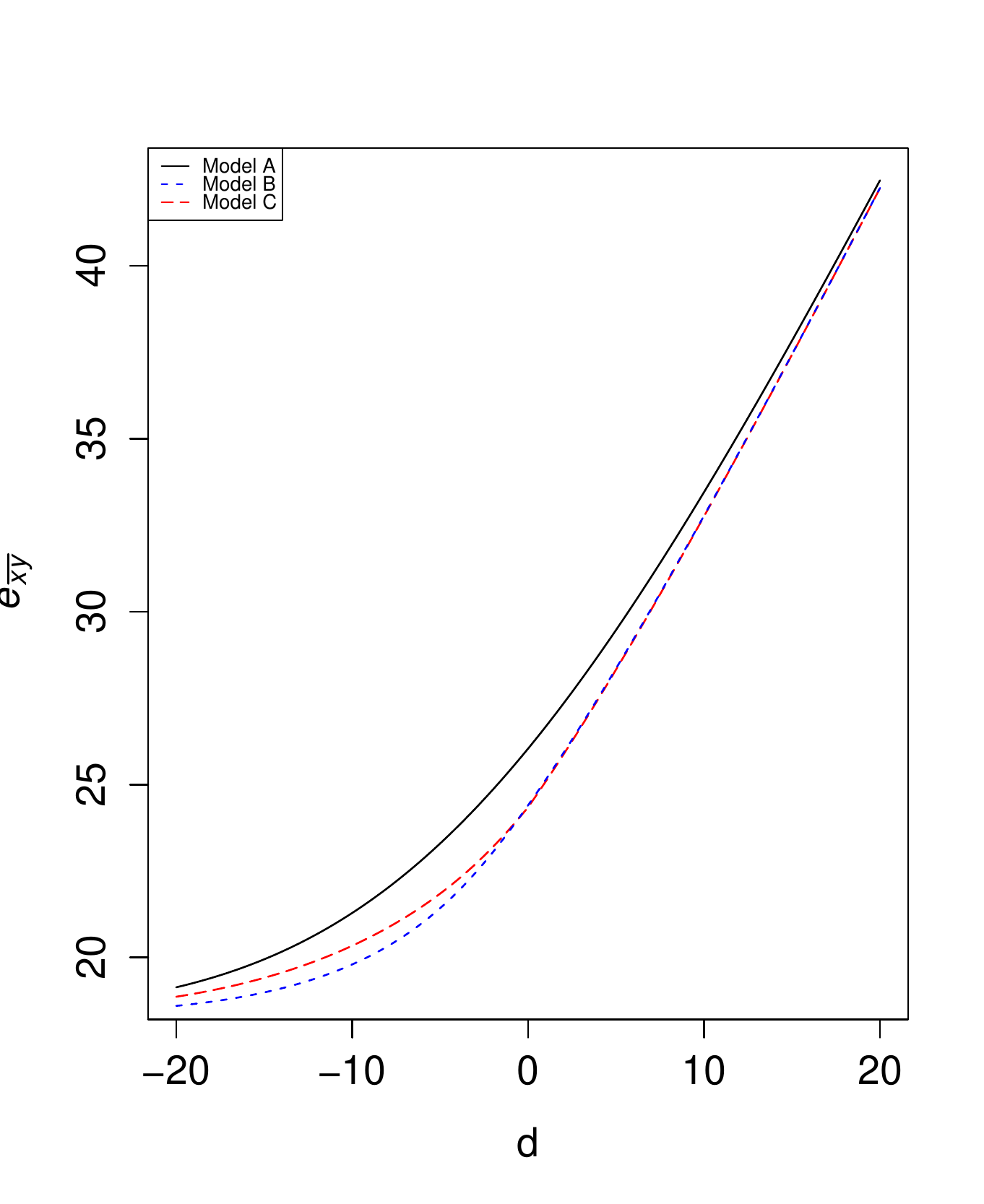}
\vspace{-0.75cm}
\label{ageddpdceratio1_10scenarios_sanstitre}
\subcaption{Joe copula: $x=65$}
\end{minipage} \hfill
\vspace{-0.35cm}
\begin{minipage}[c]{.46\linewidth}
\includegraphics[width=7.5cm,height=7.5 cm]{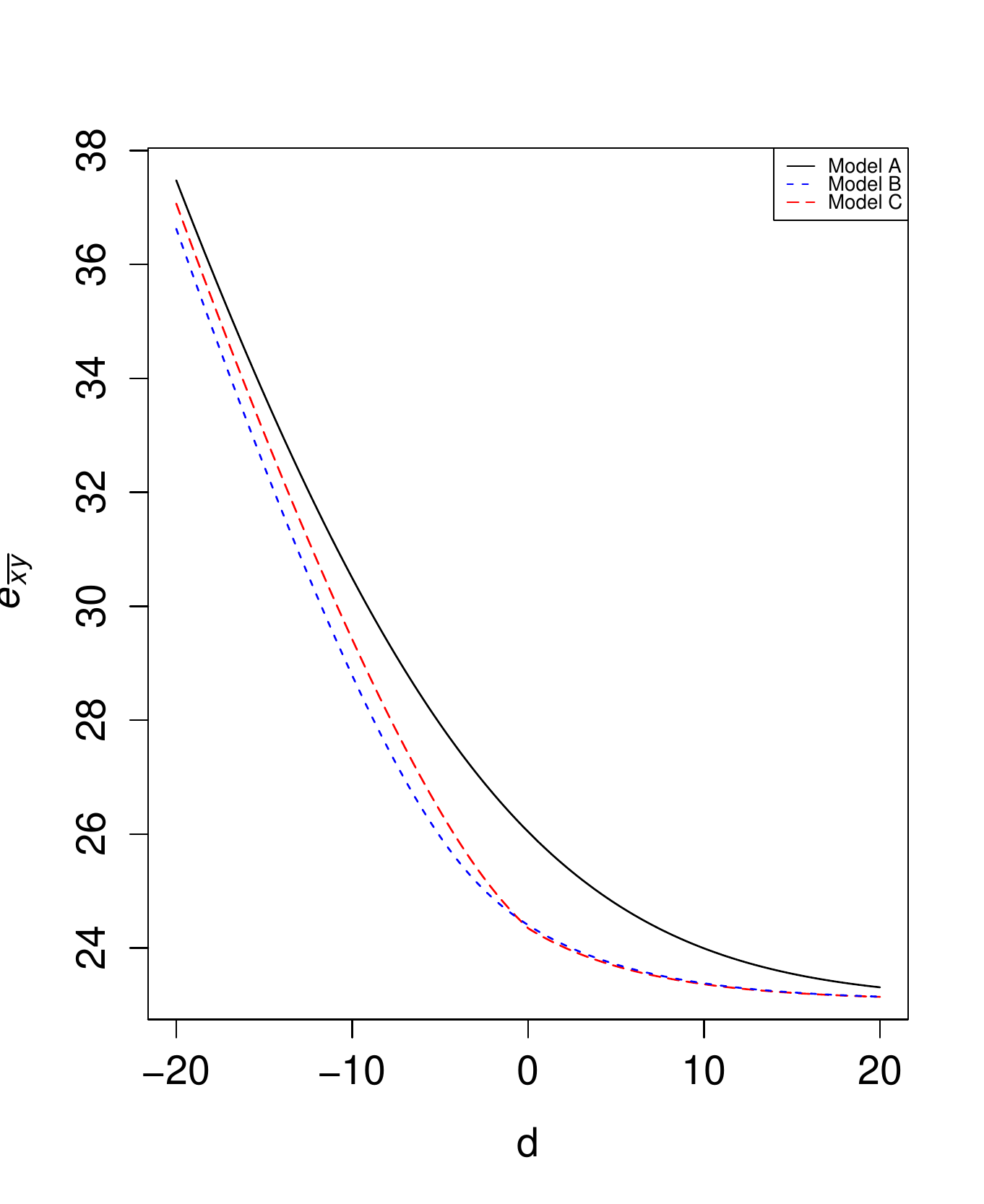}
\vspace{-0.75cm}
\label{ageddpdceratio1_10scenarios_sanstitre}
\subcaption{Joe copula: $y=65$}
\end{minipage}
\vspace{0.25cm}

\caption{Comparison of $\mathit{e}_{\overline{xy}}$ under model A, B and C: Clayton and Joe copulas}

\label{life_expectancy_comparison_1}

\end{figure}
When comparing the models A, B and C, it can be seen that the life expectancy $\mathit{e}_{\overline{xy}}$ is clearly overvalued under the model A of independence assumption, thus confirming the \red{results} obtained \red{in} \green{\cite{frees1996annuity, youn1999statistical, denuit1999multilife}}. Now, let us focus our attention on models B and C considering only Gumbel, Frank and Joe copulas as it has been shown in the previous section that the Clayton copula might not be appropriate for the Canadian insurer's data.
In all graphs, the life expectancy is always lower or equal under model B and the rate of decreases may \yT{exceed}\COM{may} $2\%$. The largest decrease is observed when $d<0$, i.e. when husband is younger than wife.
\\In order to illustrate the importance of these differences\COM{observed  in Figure \ref{life_expectancy_comparison}}, we consider four types of multiple life insurance products. Firstly, Product 1 is the \textit{joint life annuity} \red{which pays benefits} until the death of the first of the two annuitants. For a husband $\left( x\right) $ and his wife $\left( y\right) $ who receive continuously a rate of $1$, the present value of future obligations and its expectation are given by
\begin{equation}
\bar{a}_{\halfbox{T\left( xy\right)}}=\frac{1-\exp \left(  -\delta T\left( xy\right)\right) }{\delta} \;\;\;\;\;\;  \text{     and     } \;\;\;\;\;\;   \bar{a}_{xy}=\mathbb{E}\left( \bar{a}_{\halfbox{T\left( xy\right)}} \right)
\nonumber
\end{equation}
where $\delta$ is the constant instantaneous interest rate (also called force of interest). The variable $\bar{a}_{\halfbox{T\left( xy\right)}}$ can be seen as the insurer liability regarding $\left( xy\right) $. Product 2 is the last survivor annuity which pays a certain amount until the time of the second death $T\left( \overline{xy}\right)$. In that case, the present value of \green{ future} annuities and its expectation are given by
\begin{equation}
\bar{a}_{\halfbox{T\left( \overline{xy}\right)}}=\frac{1-\exp \left(  -\delta T\left( \overline{xy}\right)\right) }{\delta} \;\;\;\;\;\;  \text{     and     } \;\;\;\;\;\;   \bar{a}_{\overline{xy}}=\mathbb{E}\left( \bar{a}_{\halfbox{T\left( \overline{xy}\right)}} \right)
\nonumber
\end{equation}
In practice, payments often start at a higher level when both beneficiaries are alive. It drops at a lower level on the death of either and continues until the death of the survivor. This case is emphasized by product 3 where the rate is 1 when both annuitant are alive and reduces to $\frac{2}{3}$ after the first death. Product 3 is actually a combination of the two first annuities. Thus, the insurer \green{ liabilities} and its expectation are given by
\begin{equation}
V\left( \overline{xy}\right) =\frac{1}{3}\bar{a}_{\halfbox{T\left( xy\right)}}+\frac{2}{3}\bar{a}_{\halfbox{T\left( \overline{xy}\right)}} \;\;\;\;\; \text{     and     } \;\;\;\;\; \mathbb{E}\left( V\left( \overline{xy}\right)\right) =V_{\overline{xy}}=\frac{1}{3}\bar{a}_{xy} + \frac{2}{3}\bar{a}_{\overline{xy}}
\nonumber
\end{equation}
where $\mathbb{E}\left( \bar{a}_{\halfbox{T\left( \overline{xy}\right)}} \right)= \bar{a}_{\overline{xy}}$.
\\Fourthly, imagine a family or couple whose income is mainly funded by the husband. The family may want to guarantee its source of income for the eventual death of the husband. For this purpose, the couple may buy the so called \textit{reversionary annuity} for which the payments start right after the death of $\left( x\right) $ until the death of $\left( y\right) $. No payment is made if $\left( y\right) $ dies before $\left( x\right) $. As for Product 3, the reversionary annuity (Product 4) is also  a combination of some specific annuity policies and the total \green{obligations} \red{of the insurer} and its expectation are computed \green{as follows}
\begin{equation}
\bar{a}_\halfbox{T\left( x\right)|T\left( y\right)}= \bar{a}_{\halfbox{T\left( y\right)}} - \bar{a}_{\halfbox{T\left( xy\right)}} \;\;\;\;\; \text{     and     }  \;\;\;\;\; \bar{a}_{x|y} =  \mathbb{E}\left(  \bar{a}_\halfbox{T\left( x\right)|T\left( y\right)} \right)  =\bar{a}_y - \bar{a}_{xy}.
\end{equation}
In what follows, considering each of the insurance products 1, 2, 3 and 4, \green{comparison}  of models A, B and C will be discussed. The analysis will include the valuation of the best estimate (BE) of the aggregate liability of the insurer as well as the quantification of  risk capital and \red{stop loss premiums}.

\subsection{\eH{Risk Capital \& Stop-Loss Premium}}\label{subsec: res}
\red{In the enterprise risk management framework}, insurers are required to hold a certain capital. This amount, known as the \textit{risk capital}, \green{is used} as a buffer against unexpected large losses. The value of this capital is quantified in a way that the insurer is able to cover its liabilities  with a high probability. For instance, under Solvency II, 
it is the \textit{Value-at-Risk}(VaR) at a tolerance level of $99.5\%$ of the insurer total liability, while for the Swiss Solvency Test (SST), it is the \textit{Expected Shortfall} (ES) at $99 \%$. Let  $\zE{L} $ be the aggregate liability of the insurer. At a confidence level $\alpha$, the VaR is given by 
\BQNY
			VaR_{\zE{L}}(\alpha) =\inf \left\lbrace l\in \mathbb{R}: \mathbb{P}\left( \zE{L} \leq l\right) \geq \alpha\right\rbrace,
\EQNY
\green{whilst} the ES is
\begin{equation}
ES_{\zE{L}}(\alpha)=\mathbb{E}\left( \zE{L}|\zE{L} > VaR_{\zE{L}} (\alpha)\right).
\nonumber
\end{equation}
These risk measures will serve to compare models A, B and C for each type of product. As the insurance portfolio is made of $n$ policyholders, we define 
$$\zE{L} = \sum_{i\zE{=1}}^n L_i,$$  
where $L_i$ represents the total amount due to a couple $i$ of $\left( x_i\right) $ and $\left(y_i\right) $. The dataset used in the calculations is the same as those used for the model estimations and described in Section \ref{sec:motivation}. In principle, the couple $i$ receives the amount $b_i$ at the beginning of each year until the death of the last survivor. However, in our applications, $b_i$ will be the continuous benefit rate in CAD for each type of product. For example, in the particular case of Product 3,
\BQNY
		L_i=b_i V\left( \overline{x_i y_i}\right) =b_i \left(\frac{1}{3} \bar {a}_{\halfbox{T\left( x_i,y_i \right)}} + \frac{2}{3}\bar {a}_{\halfbox{T\left( \overline{x_i,y_i} \right)}} \right).
\EQNY
Since there is no explicit form for the distribution of $\zE{L}$, a simulation approach  will serve to evaluate the insurer aggregate liability. The pseudo-algorithm used for simulations is presented in the following steps: 
		\BIT
		 \item \textbf{Step 1}  For each couple $i$, generate $(U_i, V_i)$ from the the copula model (model A or model B or model C).
		 \item \textbf{Step 2}  For each couple $i$ with $x_i$ and $y_i$, generate the future lifetime  $T(x_i), T(y_i) $ from the Gompertz distribution \red{as follows}
\begin{equation}
T(x_i) = F_{x_i}^{-1}( U_i,\hat{\theta}_m ) \;\;\;\;\;\; \text{   and   }  \;\;\;\;\;\; T(y_i) = F_{y_i}^{-1}( V_i,\hat{\theta}_f ),
\end{equation}
where  $\hat{\theta}_j,\; j=m,f$ are taken from Table \ref{table: Gompertz_estimates}.
		 \item \textbf{Step 3}  Evaluate the liability $L_i$ for each couple $ i= 1,\ldots,n$.
		 \item \textbf{Step 4}  Evaluate the aggregate  liability of the insurer $\zE{L} = \sum_{i=1}^nL_i $. 
		 \EIT
Due to its goodness of fit performance, the Gumbel copula will be used in the calculations for Model B and C. Mortality risk is \red{assumed to be}  the only source of uncertainty and  we consider a constant  force of interest of  $\delta=5 \%$. 
For each product described in Subsection \ref{subSec:defProb}, Step 1-4 are repeated 1000 times in order to generate the distribution of $\zE{L}$. In addition to the risk capital measured as under the Solvency II and the SST framework, the $BE$ of the aggregate liability of the insurer  (i.e. $BE= \mathbb{E}\left( \zE{L} \right) $), the Coefficient of Variation (CoV)  and the Stop-Loss premium $SL= \mathbb{E}((\zE{L} - \zeta )_+)$   are also evaluated, where $\zeta$ is the deductible. 
For the portfolio of Product 1, Product 2, Product 3 and Product 4\yT{,} the amount of  $\zeta$  \green{in  millions CAD are respectively $4,$ $4.5,$ $4.2,$ $1.7.$}
Results are presented in  Table $\ref{table:Portfolioh}-\ref{table:PortfolioC}$ according to each product.
For the ease of understanding  all values have been converted to a per Model A basis (the corresponding amounts are presented in Appendix \ref{app:RM}). As we could expect, the Model A with independent lifetime assumption misjudges the total liability of the insurer. The highest differences are observable with Product 4 where it reaches $20\%$ for the $BE$, $30\%$ for the risk capitals and $71\%$ for the stop loss premiums. By comparing Model B and Model C, the findings tell  minor differences. The variation noticed in Figure \ref{life_expectancy_comparison} (when $d<0$) are practically non-existent in the aggregate values for most of the products under investigation. In other words, while the effects of the  age difference and its sign are noticeable on the individual liability (see Subsection \ref{subSec:defProb}), the  effects on the aggregate liability are merely small. \red{This} is due to the law of large number and to the high proportion  of couple with $d>0$ in our portfolio ($70\%$).  \green{Actually,  the compensation of the positive and negative effects of  the  age difference on the lifetimes dependency in the whole portfolio  mitigates its effects on the aggregate liability}. However, it should be noted that the relative difference exceeds $1.4 \%$ for the $VaR_{\zE{L}}(0.95)$ in Table \ref{table:PortfolioC}.

\begin{center}	
		\begin{tabular}{|c|c|c|c| c| c| }
		\hline 
Product 1  & BE & CoV & SL     & $VaR_{\zE{L}}(99.5\%) $ & $ES_{\zE{L}}(99\%)  $         \\   
		\hline
	Model A & 1.0000 & 0.6497 &  1.0000 & 1.0000 & 1.0000 \\
	\hline 
	Model B  & 1.0708  & 0.6279	    & 1.4072 & 1.0235 &  1.0223    \\
	\hline 
	Model C 	&1.0721	& 0.6276	& 	1.4157	 & 1.0240	 & 1.0228\\
	\hline 
		\end{tabular}
		\captionof{table}{ Relative BE and risk capital for the joint life annuity portfolio.} \label{table:Portfolioh}
\end{center}
\begin{center}	
		\begin{tabular}{|c|c|c|c| c| c|c|  c|}
		\hline 	
Product 2  &  BE & CoV & SL     & $VaR_{\zE{L}}(99.5\%) $ & $ES_{\zE{L}}(99\%)  $         \\    
		\hline
	Model A & 1.0000 &  0.5039  & 1.0000 & 1.0000 & 1.0000 \\
	\hline 
	Model B	& 0.9518 & 	0.5251	  & 	0.9220  & 0.9988	 & 0.9991 \\
	\hline 
	Model C & 0.9510	& 0.5257	 &  0.9204 	  & 	0.9989 &	0.9991    \\
	\hline 
		\end{tabular}
		\captionof{table}{ Relative BE and risk capital for the last survivor annuity (Product 2) portfolio.} \label{table:PortfolioA}
\end{center}
\begin{center}	
		\begin{tabular}{|c|c|c|c| c| c|c|  c|}
		\hline 
Product 3  &  BE & CoV & SL     & $VaR_{\zE{L}}(99.5\%) $ & $ES_{\zE{L}}(99\%)  $         \\     
		\hline
	Model A & 1.0000 &  0.5039  & 1.0000 & 1.0000 & 1.0000 \\
	\hline 
	Model B  & 0.9820  & 	0.5425  	 & 1.2148  &	1.0154	& 1.0146  \\
	\hline 
	Model C & 0.9818 &	0.5431   &	1.2191  &	1.0159 & 	1.0150    \\
	\hline 
		\end{tabular}
		\captionof{table}{ Relative BE and risk capital for the last survivor annuity (Product 3) portfolio.} \label{table:PortfolioB}
\end{center}
\begin{center}	
		\begin{tabular}{|c|c|c|c| c| c|c|  c|}
				\hline 
Product 4  &  BE & CoV & SL     & $VaR_{\zE{L}}(99.5\%) $ & $ES_{\zE{L}}(99\%)  $         \\   
		\hline
		Model A & 1.0000 &  0.5039  & 1.0000 & 1.0000 & 1.0000 \\	
	\hline 
	Model B  & 0.8072 &	1.0692	  &  0.2877 &	0.7077  &	0.7222    \\
	\hline 
	Model C & 0.8039	& 1.0586  	 &  0.2731	 & 0.6978	 &0.7135  \\
	\hline 
		\end{tabular}
		\captionof{table}{ Relative BE and risk capital for the contingent annuity portfolio.} \label{table:PortfolioC}
\end{center}

 \section{Conclusion} \label{sec: conclusion}
In this paper, we propose both parametric and semi-parametric techniques to model bivariate lifetimes commonly seen in the joint life insurance practice. The dependence factors between lifetimes are examined namely the age difference between spouses and the \eH{gender} of the elder partner in the couple. Using real insurance data, we develop an appropriate estimator of the joint distribution of the lifetimes of spouses with copula models in which the  association parameters have been allowed to incorporate the aforementioned  dependence factors. A goodness of fit procedure clearly shows that the introduced models outperform the models without age factors. The results of our illustrations, focusing on valuation of joint life insurance products, suggest that lifetimes dependence factors should be taken into account  when evaluating the best estimate of the annuity products   involving spouses.

\appendixtitleon
\appendixtitletocon
\begin{appendices}
\section{Risk measures for the aggregate liability of the insurer.} \label{app:RM}

\begin{center}	
		\begin{tabular}{|c|c|c|c| c| c|c|  }
			\hline 
Product 1  &  Mean & CoV & SL     & $VaR_{\zE{L}}(99.5\%) $ & $ES_{\zE{L}}(99\%)  $         \\      
		\hline
	Model A & 1'815'490	& 0.649  &	 31'393    &	 5'031'430 	& 5'083'090 \\
	\hline 
	Model B  & 1'944'105  & 0.628    &   44'177  &  5'149'873  &  5'196'529    \\
	\hline 
	Model C & 1'946'400 & 0.627   &    44'443   &  5'152'233  &  5'199'015   \\
	\hline 
		\end{tabular}
		\captionof{table}{ Risk capital for the joint life annuity portfolio in CAD.} \label{table:jointlifePortfolio1}
\end{center}
\begin{center}	
		\begin{tabular}{|c|c|c|c| c| c|c|  c|c|}
		\hline 
Product 2  &  Mean & CoV & SL     & $VaR_{\zE{L}}(99.5\%) $ & $ES_{\zE{L}}(99\%)  $         \\    
		\hline
	Model A & 2'663'056 & 0.487    &   61'826   &  5'557'880  &  5'590'822\\
	\hline 
	Model B  & 2'534'628  & 0.525    &   57'007    &  5'551'368   &  5'585'636   \\
	\hline 
	Model C & 2'532'504  & 0.526   &   56'906  &  5'551'814  & 5'585'818     \\
	\hline 
		\end{tabular}
		\captionof{table}{ Risk capital for the last survivor annuity (Product 2) portfolio in CAD.} \label{table:jointlifePortfolio2}
\end{center}
\begin{center}	
		\begin{tabular}{|c|c|c|c| c| c|c|  c|}
		\hline 
Product 3  &  Mean & CoV & SL     & $VaR_{\zE{L}}(99.5\%) $ & $ES_{\zE{L}}(99\%)  $         \\    
		\hline
	Model A & 2'380'534  & 0.504   &  50'205    &  5'275'035  & 5'316'415 \\
	\hline 
	Model B  & 2'337'787  & 0.543      &  60'990     &  5'356'069  &  5'394'256   \\
	\hline 
	Model C & 2'337'136  & 0.543    &   61'206   &  5'358'722  &  5'396'062    \\
	\hline 
		\end{tabular}
		\captionof{table}{ Risk capital for the  last survivor  annuity (Product 3)					 portfolio in CAD.} \label{table:jointlifePortfolio3}
\end{center}
\begin{center}	
		\begin{tabular}{|c|c|c|c| c| c|c| c|}
		\hline 
Product 4  &  Mean & CoV & SL     & $VaR_{\zE{L}}(99.5\%) $ & $ES_{\zE{L}}(99\%)  $         \\   
		\hline
	Model A & 667'479  &  1.248    &   93'413  &  4'123'250  &  4'200'646 \\
	\hline 
	Model B  & 538'811  &  1.069     &  26'871    &  2'918'125  &  3'033'624    \\
	\hline 
	Model C & 536'592  & 1.059     &   25'514   &  2'877'347  &   2'997'130   \\
	\hline 
		\end{tabular}
		\captionof{table}{ Risk capital for the life contingent annuity  portfolio in CAD.} \label{table:jointlifePortfolio4}
\end{center}

\end {appendices}


\textbf{Acknowledgments}
\\
The authors acknowledge partial support from a Swiss National Science Foundation grant and the project RARE -318984
 (an FP7  Marie Curie IRSES Fellowship). Gildas Ratovomirija is partially supported by  Vaudoise Assurances.
The authors would also like to thank Nicolas Salani for the interesting discussions during the preparation of this contribution.

\bibliographystyle{plain}
\bibliography{BG}

\begin{thebibliography}{10}

\bibitem{CoriA}
H.~Albrecher, C.~Constantinescu, and S.~Loisel.
\newblock Explicit ruin formulas for models with dependence among risks.
\newblock {\em Insurance Math. Econom.}, 48(2):265--270, 2011.

\bibitem{berg2009copula}
D.~Berg.
\newblock Copula goodness-of-fit testing: an overview and power comparison.
\newblock {\em The European Journal of Finance}, 15(7-8):675--701, 2009.

\bibitem{bowers1986actuarial}
N.L. Bowers, H.U. Gerber, J.C. Hickman, D.A. Jones, and C.J. Nesbitt.
\newblock {\em Actuarial mathematics}, volume~2.
\newblock Society of Actuaries Itasca, Ill., 1986.

\bibitem{brown1999joint}
J.R. Brown and J.M. Poterba.
\newblock Joint life annuities and annuity demand by married couples.
\newblock Technical report, National bureau of economic research, 1999.

\bibitem{carriere1994investigation}
J.F. Carriere.
\newblock An investigation of the {G}ompertz law of mortality.
\newblock {\em Actuarial Research Clearing House}, 2:161--177, 1994.

\bibitem{carriere2000bivariate}
J.F. Carriere.
\newblock Bivariate survival models for coupled lives.
\newblock {\em Scandinavian Actuarial Journal}, 2000(1):17--32, 2000.

\bibitem{CoriB}
C.~Constantinescu, E.~Hashorva, and L.~Ji.
\newblock Archimedean copulas in finite and infinite dimensions---with
  application to ruin problems.
\newblock {\em Insurance Math. Econom.}, 49(3):487--495, 2011.

\bibitem{dabrowska1988kaplan}
D.M. Dabrowska.
\newblock Kaplan-meier estimate on the plane.
\newblock {\em The Annals of Statistics}, pages 1475--1489, 1988.

\bibitem{denuit1999multilife}
M.~Denuit and A.~Cornet.
\newblock Multilife premium calculation with dependent future lifetimes.
\newblock {\em Journal of Actuarial Practice}, 7:147--171, 1999.

\bibitem{denuit2001measuring}
M.~Denuit, J.~Dhaene, C.~Le~Bailly~de Tilleghem, and S.~Teghem.
\newblock Measuring the impact of dependence among insured lifelengths.
\newblock {\em Belgian Actuarial Bulletin}, 1(1):18--39, 2001.

\bibitem{frees1996annuity}
E.W. Frees, J.F. Carriere, and E.~Valdez.
\newblock Annuity valuation with dependent mortality.
\newblock {\em Journal of Risk and Insurance}, pages 229--261, 1996.

\bibitem{genest1995semiparametric}
C.~Genest, K.~Ghoudi, and L.P. Rivest.
\newblock A semiparametric estimation procedure of dependence parameters in
  multivariate families of distributions.
\newblock {\em Biometrika}, 82(3):543--552, 1995.

\bibitem{genest2009goodness}
C.~Genest, B.~R{\'e}millard, and D.~Beaudoin.
\newblock Goodness-of-fit tests for copulas: A review and a power study.
\newblock {\em Insurance: Mathematics and economics}, 44(2):199--213, 2009.

\bibitem{gribkova2015non}
S.~Gribkova and O.~Lopez.
\newblock Non-parametric copula estimation under bivariate censoring.
\newblock {\em Scandinavian Journal of Statistics}, 2015.

\bibitem{joe1996estimation}
H.~Joe and J.J. Xu.
\newblock The estimation method of inference functions for margins for
  multivariate models.
\newblock Technical report, Technical report, 1996.

\bibitem{lawless2011statistical}
J.F. Lawless.
\newblock {\em Statistical models and methods for lifetime data}, volume 362.
\newblock John Wiley \& Sons, 2011.

\bibitem{luciano2008modelling}
E.~Luciano, J.~Spreeuw, and E.~Vigna.
\newblock Modelling stochastic mortality for dependent lives.
\newblock {\em Insurance: Mathematics and Economics}, 43(2):234--244, 2008.

\bibitem{maeder1996construction}
Ph. Maeder.
\newblock La construction des tables de mortalite du tarif collectif 1995 de
  l'{UPAV}.
\newblock {\em Insurance Mathematics and Economics}, 3(18):226, 1996.

\bibitem{nelsen2007introduction}
R.B. Nelsen.
\newblock {\em An introduction to copulas}.
\newblock Springer Science \& Business Media, 2007.

\bibitem{oakes1989bivariate}
D.~Oakes.
\newblock Bivariate survival models induced by frailties.
\newblock {\em Journal of the American Statistical Association},
  84(406):487--493, 1989.

\bibitem{parkes1970broken}
C.M. Parkes, B.~Benjamin, and R.G. Fitzgerald.
\newblock Broken heart: A statistical study of increased mortality among
  widowers.
\newblock {\em Journal of Occupational and Environmental Medicine}, 12(4):143,
  1970.

\bibitem{prentice2004hazard}
R.L. Prentice, F.~Zoe~Moodie, and J.~Wu.
\newblock Hazard-based nonparametric survivor function estimation.
\newblock {\em Journal of the Royal Statistical Society: Series B (Statistical
  Methodology)}, 66(2):305--319, 2004.

\bibitem{shih1995inferences}
J.H. Shih and T.A. Louis.
\newblock Inferences on the association parameter in copula models for
  bivariate survival data.
\newblock {\em Biometrics}, pages 1384--1399, 1995.

\bibitem{sklar1959fonctions}
M.~Sklar.
\newblock {\em Fonctions de r{\'e}partition {\`a} n dimensions et leurs
  marges}.
\newblock Universit{\'e} Paris 8, 1959.

\bibitem{ward1976mortality}
A.W. Ward.
\newblock Mortality of bereavement.
\newblock {\em BMJ}, 1(6011):700--702, 1976.

\bibitem{youn1999statistical}
H.~Youn and A.~Shemyakin.
\newblock Statistical aspects of joint life insurance pricing.
\newblock {\em 1999 Proceedings of the Business and Statistics Section of the
  American Statistical Association}, 34:38, 1999.

\end{thebibliography}
\end{document}